\documentclass[11pt,preprintnumbers,nofootinbib,prd,axodraw]{revtex4}

\usepackage{graphicx}
\usepackage{amsmath}
\usepackage{epsfig}
\usepackage{mathrsfs}

\newcommand{\be}{\begin{equation}}
\newcommand{\ee}{\end{equation}}
\newcommand{\ba}{\begin{eqnarray}}
\newcommand{\ea}{\end{eqnarray}}
\newcommand{\notE}{E\kern-0.6em\hbox{/}\kern0.05em}
\newcommand{\notEt}{E_{T}\kern-1.21em\hbox{/}\kern0.45em}
\newcommand{\notsusy}{SUSY\kern-1.21em\hbox{/}\kern0.45em}

\def\us{\char`\_}
\def\lsim{\mathrel{\rlap{\lower4pt\hbox{\hskip1pt$\sim$}}
     \raise1pt\hbox{$<$}}}                
\def\gsim{\mathrel{\rlap{\lower4pt\hbox{\hskip1pt$\sim$}}
     \raise1pt\hbox{$>$}}}                

\begin{document}
\begin{flushright}
UCB-PTH-09/04\\
\end{flushright}
\title{The Leptonic Higgs as a Messenger of Dark Matter}
\author{Hock-Seng Goh$^1$}
\author{Lawrence J. Hall$^{1,2}$}
\author{Piyush Kumar$^1$}
\affiliation{$^1$Berkeley Center for Theoretical Physics, University of California,
Berkeley, CA 94720\;\;$\&$\\Theoretical Physics Group, Lawrence Berkeley National Laboratory, Berkeley, CA 94720\\
$^2$Institute for Physics and Mathematics of the Universe (IPMU)\\University of Tokyo, Kashiwa-no-ha 5-1-5, 277-8592, 
Japan.\\ \\}
\begin{abstract}
We propose that the leptonic cosmic ray signals seen by PAMELA and ATIC result from
the annihilation or decay of dark matter particles via states of a leptonic Higgs doublet to
$\tau$ leptons, linking cosmic ray signals of dark matter to LHC signals of the Higgs sector.
The states of the leptonic Higgs doublet are lighter than about 200 GeV, yielding large
$\bar{\tau} \tau$ and  $\bar{\tau} \tau \bar{\tau} \tau$ event rates at the LHC. Simple models are given for
the dark matter particle and its interactions with the leptonic Higgs, for cosmic ray
signals arising from both annihilations and decays in the galactic halo.  For the case of
annihilations, cosmic photon and neutrino signals are on the verge of discovery.
\end{abstract}
\maketitle
\newpage
\vspace{-1.2cm} \tableofcontents

\section{Introduction}

Recent observations of high-energy electron and positron cosmic ray spectra
have generated tremendous interest, as they might provide the first
non-gravitational evidence for Dark Matter (DM).
The PAMELA \cite{pamela} experiment reports an excess of positrons in
the few GeV to 100 GeV range, providing further support to the
earlier results of HEAT \cite{heat} and AMS \cite{ams01}. In
addition, results from the ATIC \cite{atic} and PPB-BETS
\cite{ppb} balloon experiments suggest an excess of electrons and
positrons in the 300 GeV to 600 GeV range.

While these observations have conventional astrophysical interpretations,
they may result from annihilations or decays of
DM particles in the galactic halo. Indeed,
the PAMELA and ATIC data reinforce each other, since, for
a certain range of DM masses, they have a unified interpretation.
However, DM explanations for the leptonic cosmic ray excesses face
two interesting challenges. First, for annihilating DM these signals require that the
annihilation cross-section for DM particles is typically two to
three orders of magnitude larger than that expected from the thermal
freezeout of WIMP DM. On the other hand, for decaying DM, the life-time of the DM particles must
be extremely large, of ${\cal O}(10^{25-26})$ seconds. Second, the signals apparently require
annihilations or decays dominantly into leptons
rather than hadrons, since there is no reported excess in
anti-proton cosmic rays. Inspite of these challenges, many papers
with different models of DM have already appeared in the
literature, utilizing both annihilations \cite{kin-lepton,Cholis:2008wq,Harnik:2008uu,weiner-L}
and decays \cite{decay,Ibarra:2008jk,Nardi:2008ix}.

For annihilating DM, many of these models try to explain the required
large annihilation cross-section by a Sommerfeld enhancement \cite{Hisano}, which
is operative at the non-relativistic velocities ($\beta \sim
10^{-3}$) in the galactic halo, while still having a standard
thermal relic abundance applicable at the time of DM freeze-out.
There are several possibilities for understanding why the products of the annihilation
are dominantly leptonic rather than hadronic.  One possibility is kinematics:
the annihilation products of the DM particles are not heavy enough to decay
into quarks, gauge bosons, and Higgs bosons, which have a large
hadronic branching ratio, and hence decay only to electrons and muons (and possibly taus).

Another possibility is a \emph{symmetry}, rather than kinematics, to understand why the
DM annihilation or decay products are dominantly leptonic.  In this work, we study a DM sector
coupling to the visible sector through Higgs messengers which couple to leptons due to a symmetry. 
The plan of the paper is as follows. In the next section, we motivate this possibility 
from a general perspective, stressing that
this is a natural implementation of the hypothesis that the DM is a WIMP, with mass and
interactions broadly governed by the mass scale of weak interactions.
In section \ref{leptonichiggs}, we give the yukawa interactions for the minimal leptonic Higgs theory,
and write them in a mass eigenstate basis for both the Higgs and the quarks and leptons.
There are two Higgs scalars, $h$ and $H$, one pseudoscalar, $A$, and one charged Higgs boson $H^+$,
each with matter interactions that are determined by the ratio of vevs $\tan \beta$
and the Higgs mixing angle $\alpha$.  These interactions are quite unlike those of the usually
considered two Higgs doublet model or of the MSSM.  The LEP mass limits for $h,H,A$ and $H^+$ are given, as well
as constraints that follow from the cosmic ray signals.

In section \ref{astro} the possible classes of DM annihilations and decays through the leptonic Higgs
states $H,A$ and $H^+$ are discussed.  The common feature is multi-$\tau$ final states.  For a particular
annihilation channel the cosmic-ray electron and positrons signals are derived and compared to the
PAMELA and ATIC data.  In general, if the PAMELA and ATIC data result from enhanced galactic DM
annihilation, then significant fluxes of photons \cite{Bell:2008vx,Bertone:2008xr,Mardon:2009rc} and neutrinos \cite{Hisano:2008ah,Liu:2008ci} are also expected.
On the one hand this could allow for a crucial confirmation of the DM nature of the signal, while on the
other hand there is frequently some tension with present limits on photon and neutrino fluxes.  We calculate
photon fluxes in the leptonic Higgs model from DM annihilations in both the galactic center
of the Milky Way and in the dwarf galaxy Sagittarius, and also discuss the neutrino flux.

In section \ref{higgs-lhc}, we study the implications for Higgs signals at the LHC.
The LHC Higgs signals are  \emph{necessarily correlated} with the cosmic-ray signals,
with the leptonic Higgs states $H, A$ and $H^+$ decaying dominantly to final states
involving the $\tau$ lepton.  The collider phenomenology is very rich, and quite
unlike that of conventional two Higgs doublet models, such as the MSSM.
In section \ref{explicit}, we study some simple models for the DM particle and its
couplings to the Higgs sector, including the cases that the DM particle is derived
from electroweak singlet and doublet fermions or scalars.  Since our dark matter particle
is heavy, in the few TeV region, for DM annihilations the coupling strength to the leptonic
Higgs is typically quite strong. In addition, a Sommerfeld enhancement of the annihilation
cross section typically results from the exchange of the leptonic Higgs, which is lighter
than $2 M_W$.  In the case of DM decays, the long lifetime results partly from the symmetry
that forces one Higgs to be leptonic.
We conclude in section \ref{conclude}.

\section{The Higgs as a Messenger of Dark Matter Signals}\label{messenger}

\begin{figure}[h!]
\begin{center}
\resizebox{3.5in}{!}{\includegraphics*[210,625][470,690]{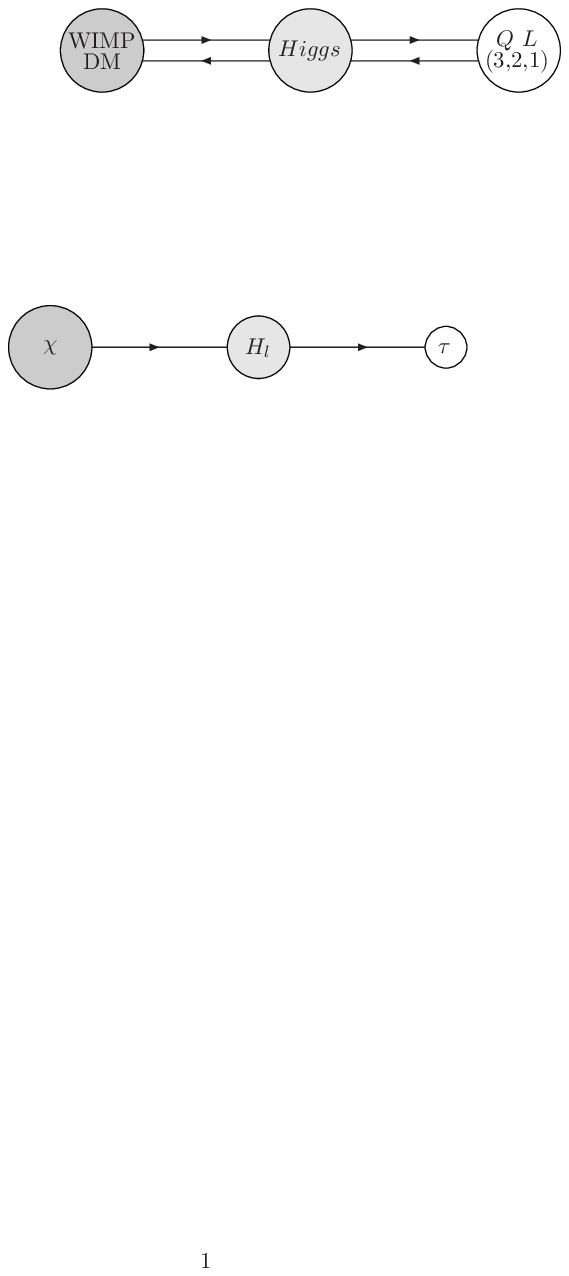}}
\end{center}
\caption{Three sectors and their interactions:  the known quarks
and leptons and their $SU(3)\times SU(2)\times U(1)$ gauge
interactions, a Higgs sector and a WIMP Dark Matter sector.}
\label{cartoon}
\end{figure}
What is our best guess for the structure of particle interactions
at the TeV scale? In addition to the known physics of quarks and
leptons interacting via $SU(3) \times SU(2) \times U(1)$ gauge
interactions, we expect new physics to include both a Higgs
sector, responsible for $SU(2) \times U(1)$ symmetry breaking, and
a dark matter sector. While dark matter need not be related to the
TeV scale, the WIMP idea is intriguing: if the dark matter
particle mass is of order the TeV scale, the order of magnitude of
the observed abundance results from thermal freezeout using simple
dimensional analysis.  Thus we are led to the three sectors of
Figure 1: the known sector of quarks and leptons and their gauge
interactions, together with the unknown sectors of the Higgs and
WIMP dark matter.

What are the interactions between these sectors?  Clearly the
Higgs interacts directly with the quarks and leptons and $SU(2)
\times U(1)$ gauge interactions, as shown, to give the observed
masses.  The interactions of the WIMP dark matter sector with the
other two sectors are more speculative.  In fact, there need not
be any; the WIMP could annihilate to extremely light particles in
its own sector. However, the WIMP idea is that the mass scales of
the dark matter and Higgs sectors are related, and this strongly
suggests some connection between these two sectors, as shown in
Figure 1. Indeed, it could be that the Higgs and dark matter
sectors are so closely connected that they merge. We assume that
any direct couplings between the WIMP sector and the quark and
lepton sector are subdominant, so that we are led to the pattern
of connections of Figure 1.  The Higgs sector is seen to be the
messenger that makes the WIMP visible to us.  The implications of
this, for cosmic ray signatures of dark matter is clear. 
The WIMP, $\chi$, will be observed via annihilations or decays through
the Higgs particles. A simple realization of this idea is to introduce 
a singlet scalar to the SM, since the only gauge invariant renormalizable operator 
is a coupling to the Higgs \cite{Silveira:1985rk}. This can be extended to supersymmetric theories by adding gauge  
singlet dark matter to the NMSSM \cite{MarchRussell:2008yu}.

What are the implications of the cosmic ray signals for this
picture? The exciting thing is that the
cosmic ray signals may shed light on the structure of the Higgs
sector. The signals will depend on the nature and interactions of
the Higgs states. The ATIC data suggests that the WIMP, $\chi$,
will be too heavy to be observed at the LHC, so that the collider
signals that could demonstrate consistency of the picture are
those of the Higgs sector.

Could the Higgs sector simply be the single Higgs doublet of the
Standard Model? Depending on its mass, the Higgs would decay
dominantly to pairs of top quarks, $W$ bosons, or bottom quarks.
In either case, for any mass of the WIMP, the signal in $e^+ +
e^-$ is smooth, and does not give the peak shown by the balloon
experiments \cite{Cirelli:2008pk}.  Hence the Standard Model Higgs can
lead to the PAMELA positron signal, but not the ATIC $e^+
+ e^-$ peak. If the latter is ignored, there is still the issue of
whether dominant hadronic decays will produce a $\bar{p}$ flux
larger than seen by PAMELA.  For $m_\chi < 1$ TeV the $\bar{p}$
flux is apparenty one to two orders of magnitude too large. However,
the uncertainties in the $\bar{p}$ signal are certainly an order
of magnitude \cite{Donato} and, by going to larger values of
$m_\chi$, the $\bar{p}$ spectrum can be shifted to larger energies
where there is no data. For decays to top or bottom quark pairs,
$m_\chi > 1$ TeV is in any case needed to explain the PAMELA
signal.  Values of $m_\chi$ around 100-200 GeV are possible
for decays to $W$ pairs, but this leads to some tension with data from
anti-protons and gamma rays and also requires that the
energy loss rate for positrons be larger \cite{Cirelli:2008pk,Grajek:2008pg}.


In this paper we concentrate on explanations of both the PAMELA
$e^+$ data and the ATIC $e^+ + e^-$ peak.  Quite generally this
requires dominant cascades directly to charged leptons
\cite{Cirelli:2008pk}, so that we are immediately led to a Higgs
that couples predominantly to leptons. Such Higgs bosons have
received very little attention since they are not immediate
consequences of either supersymmetric or grand unified theories
(although this is possible with some model-building). For earlier work on Higgs bosons 
coupling to leptons with different motivation, see \cite{leptonic2}.
However, a leptonic higgs is the most straightforward implication of assuming
that the cosmic ray data is explained by the annihilation or decay of WIMPs, $\chi$, through
Higgs messengers, as illustrated in Figure 1.  Furthermore, the
absence, to very high accuracy, of flavor violation in the charged
lepton sector suggests that there is a single leptonic Higgs
doublet, $H_l$.  Thus, the experimental consequences for cosmic ray observations 
result from the connections below,  
\begin{figure}[h!]
\begin{center}
\resizebox{2.5in}{!}{\includegraphics*[185,480][425,540]{dark.eps}}
\end{center} \label{eq:chainH_l}
\end{figure}
while those for the LHC involve production of leptonic Higgs states 
followed by their decay to tau leptons. 

The ATIC peak in the $e^+ + e^-$ channel will be
broader than in theories where the signal results from $\chi$
cascading directly to $e$ or $\mu$.  Also, there is the
possibility of an energetic gamma-ray and neutrino component of
the cosmic rays component in the energy range 100-1000 GeV. Both
these signals could be detected in future experiments like
GLAST/FERMI, VERITAS4 (gamma rays) and Hyper-Kamiokande, ANTARES,
KM3neT, ICE-CUBE (neutrinos). These latter signals depend
crucially on whether the cascade above results from the
annihilation or decay of halo dark matter. We consider both
possibilities in this paper.

Consider first the case that the cosmic ray signals arise from the
annihilation of dark matter in the halo to $H_l$ states. The size
of the signal requires a galactic annihilation cross section that
is significantly larger than the annihilation cross section
required for a successful thermal freezeout abundance, so that a
non-thermal production mechanism \cite{randall-moroi} is necessary. In the absence of
thermal freeezeout one might wonder whether the motivation for
WIMP dark matter is lost.  Clearly the answer is no:  the cosmic
ray signals have directly measured the annihilation cross section,
and its order of magnitude is consistent with a weak scale mass --
the WIMP motivation is actually strengthened.

On the other hand, if the cosmic ray signals result from decays of
the dark matter in the halo, then the abundance of the dark matter
could be given by the conventional thermal freezeout of WIMPs.  New
physics is now needed to induce a dark matter lifetime of order
$10^{26}$ seconds.  In the limit of stability, the theory possesses
two discrete symmetries: one that prevents $H_l$ from coupling to
quarks and another that ensures the stability of $\chi$.  The
decay chain $\chi \rightarrow H_l$ involves the breaking of both
discrete symmetries. If the discrete symmetries are spontaneously
broken at the weak scale $v$, then the dimensionless symmetry
breaking parameters may be of order $v/M$, giving a $\chi$ decay
rate of order $\Gamma_\chi \sim v^5/M^4$, which is the desired
rate for $M\simeq 10^{16}$ GeV.  While this is only a very rough
order of magnitude estimate, it can be considered to be an
extension of the WIMP idea, that the dark matter particle mass and
interactions are governed by the weak scale.

In section \ref{explicit} we will introduce some explicit models for
the dark matter sector and its coupling to the leptonic Higgs
doublet. Each model can be considered in the ``annihilation mode''
or the ``decay mode'' depending on the absence or presence of the
higher dimensional interactions that induce decay.  Since these
symmetry breaking effects are extremely small they will not affect
the LHC signals of the model.  Indeed, the main effect is on the
size of the high energy photon and neutrino signals in the cosmic rays.  For
annihilations this depends on the square of the dark matter
density, while for decays it is only linear in the density.
The photon and neutrino signals have a much larger support from the
center of the galaxy, or from satellite galaxies, while the lepton
signal, because of propagation effects, has support from regions
of the halo close to us.  Thus it is likely to be the photon
and neutrino signals that distinguish annihilations from decays. Moreover, as we will
see, the neutrino signals tend to be more robust than the photon signals.



\section{The Leptonic Higgs Boson}\label{leptonichiggs}

Here we discuss the crucial features of the leptonic Higgs boson
relevant for the PAMELA data; we postpone a discussion of the
Higgs potential until section V.   Consider a two Higgs doublet
model with a symmetry that forces one Higgs doublet, $H_l$, to
couple to the charged lepton sector and another, $H_q$, to couple
to quarks
\begin{equation}
{\cal L}_{yuk} = y_u^{ij}\, Q_i\,  u_j^c H_q^\dagger +y_d^{ij}\,
Q_i\, d_j^c H_q + y_e^{ij}\, L_i\,  e_j^c H_l + h.c.
\label{eq:yuk}
\end{equation}
where $Q_i,u^c_i, d^c_i,L_i,e^c_i$ are the quark and lepton fields with
$(i,j)$ being the family index, and $y_{u,d,e}^{ij}$ are the
Yukawa coupling matrices.

We assume that the primary cosmic ray positron spectrum arises from cascade chains of the form
\begin{equation}
\chi(\chi) \rightarrow H(A) \rightarrow \bar{\tau} \tau (\bar{\tau} \tau) \rightarrow e^+
+\ldots \label{eq:chain}
\end{equation}
where $\chi$ is some neutral dark matter particle, and $H$ and $A$
are a scalar and pseudoscalar of the Higgs sector, respectively.
The chain could arise from either annihilation or decay of the
dark matter particles, and the relative strength of the chains via
$H$ and $A$ can vary.  To suppress primary cosmic ray $\bar{p}$,
$H$ and $A$ must lie dominantly in the leptonic Higgs doublet
$H_l$. Writing the neutral component of the two Higgs doublets as
\begin{equation}
H_l^0 = v_l + \frac{h_l + i a_l}{\sqrt{2}}, \hspace{0.3in} H_q^0 =
v_q + \frac{h_q + i a_q}{\sqrt{2}}, \label{eq:neutralcomp}
\end{equation}
the $H$, $A$ and $h$ states may be written as
\begin{equation}
H = \cos \alpha \, h_l + \sin \alpha \, h_q; \hspace{0.5in} A =
\cos \beta \, a_l - \sin \beta \, a_q; \hspace{0.5in} h = \sin
\alpha \, h_l - \cos \alpha \, h_q \label{eq:SandA}
\end{equation}
where
\begin{equation}
\tan \beta = \frac{v_l}{v_q} \ll 1; \hspace{1in} v_l^2+v_q^2=v^2=(174\,{\rm GeV})^2
\label{eq:tanbeta}
\end{equation}
and the mixing angle $\alpha$, that diagonalizes the Higgs mass matrix, is also taken to be small,
\begin{equation}
\sin \alpha \ll 1.
\label{eq:alpha}
\end{equation}

While $A$ is the only pseudoscalar, we call the scalar orthogonal to the higgs boson $H$
as $h$.  Since it lies dominantly in $H_q$,
and since $v_q \gg v_l$, it is the scalar most closely related to
electroweak symmetry breaking.  It has many of the properties of
the Standard Model Higgs boson, except that its couplings to
leptons are not standard.  Also, precision electroweak data do not
require that $m_h$ is close to the present experimental bound,
since the other scalar states, and possibly the dark matter
sector, may contribute to the $S$ and $T$ observables.  In
addition there is a charged scalar, $H^+$, that lies dominantly in
$H_l$.   The couplings of $H$, $A$, $h$ and $H^+$ to quarks and
leptons are proportional to the diagonal quark and lepton mass
matrices, $m_{u,d,e}$:
\begin{eqnarray}
{\cal L}_H &=& \frac{H}{\sqrt{2}v} \left( \frac{\sin \alpha}{\cos
\beta} (u m_u u^c + d m_d d^c) \, + \, \frac{\cos \alpha}{\sin
\beta}
(e m_e e^c) \right)\nonumber\\
{\cal L}_A &=& \frac{iA}{\sqrt{2}v} \left( \frac{-\sin \beta}{\cos
\beta} (u m_u u^c + d m_d d^c) \, + \, \frac{\cos \beta}{\sin
\beta}
(e m_e e^c) \right)\nonumber\\
{\cal L}_h &=& \frac{h}{\sqrt{2}v} \left( \frac{-\cos \alpha}{\cos
\beta} (u m_u u^c + d m_d d^c) \, + \, \frac{\sin \alpha}{\sin
\beta}
(e m_e e^c) \right)\nonumber\\
{\cal L}_{H^\pm} &=& \frac{H^-}{v} \left( \frac{-\sin \beta}{\cos
\beta} (u V_{CKM} m_d d^c + u^{c^\dagger} m_u d^\dagger) \, + \,
\frac{\cos \beta}{\sin \beta} (\nu_e m_e e^c) \right)
\label{eq:SAhcouplings}
\end{eqnarray}
where $V_{CKM}$ is the CKM mixing matrix of the charged current
quark interaction.

The LEP experiments have placed bounds on the scalars $H$, $A$ and
$H^+$. The cross section for the process $e^+e^- \rightarrow ZH$
is proportional to $\sin^2(\alpha - \beta)$ and depends on $m_H$.
The limits on $\sin^2(\alpha - \beta)$ are shown in Figure 2c of
\cite{Schael:2006cr} for the case that $H$ decays dominantly to
$\bar{\tau} \tau$, as we expect.  For $\sin(\alpha - \beta) = 0.3$
the $H$ mass cannot lie in the region (30 - 100) GeV, but for
$\sin(\alpha - \beta) < 0.2$ there is no limit.  More importantly,
the LEP experiments have searched for the process $e^+e^-
\rightarrow HA$ which is proportional to $\cos^2(\alpha- \beta)
\simeq 1$. In the case that both $H$ and $A$ decay dominantly to
$\bar{\tau} \tau$ as we have, the limits are shown in Figure 4d of
\cite{Schael:2006cr} and require that either $m_H + m_A < 20$ GeV,
which is strongly excluded by the width of the $Z$ boson, or
\begin{equation}
m_H + m_A \, > \, 185 \, \mbox{GeV}. \label{eq:mS+mA}
\end{equation}
%
If the charged Higgs, $H^+$, is lighter than the $tb$ and $WZ$
thresholds, it will dominantly decay to $\tau \nu_\tau$, since the
$cs$ final state has a relative supression of $\tan^4 \beta$. Thus
the limit set by the ALEPH collaboration is \cite{chargedHiggs}
\begin{equation}
 m_{H^+} \, > \, 88 \, \mbox{GeV}.
\label{eq:mH+}
\end{equation}

In addition to these LEP constraints on the higgs masses, we
require that the $H,A \, \rightarrow \, \bar{\tau} \tau$ branching
ratios are dominant and at least 0.9. This is because the dark
matter annihilation or decay chain passes through $H/A$ states and
we need to satisfy the $\bar{p}$ constraint from PAMELA. In
particular to avoid final states involving electroweak gauge
bosons we impose
\begin{equation}
|m_H-m_A| \, < \, m_Z \label{eq:H-A}
\end{equation}
in order to suppress the mode $H\rightarrow A\,Z$ ($A\rightarrow
H\,Z$), and
\begin{equation}
m_H \, < \, 2 m_W. \label{eq:H-2W}
\end{equation}
to avoid $H \, \rightarrow \, 2W$. Finally, we require
\begin{equation}
m_H < 2\,m_A \label{eq:H-2A}
\end{equation}
in order to forbid the decay $H \rightarrow AA$ since such a
cascade of $H$ would lead to a less prominant peak in the lepton
cosmic ray spectrum for ATIC. To summarize, these
constraints impose limits on $m_H$ and
$m_A$:
\begin{eqnarray}
    \frac{2m_Z}{3} \,<\,& m_A & \,<\, 2m_W+m_Z\nonumber\\
    \frac{m_Z}{2} \,<\,& m_H & \,<\, 2m_W
\end{eqnarray}
Under these conditions, $H$ and $A$ predominantly go to $\bar{\tau}
\tau$. The decay to $b\bar{b}$ is suppressed due to the
leptophilic nature of $H$ and $A$.

How small should $\sin \alpha$ and $\sin \beta$ be taken in order
that the $H$ and $A$ couplings to quarks are small enough to
sufficiently suppress the $\bar{p}$ cosmic ray flux?  For the mass
range we are considering in this paper, the dominant decays are to
$\bar{b}b$ and $\bar{\tau} \tau$. In this case, the ratio of
quarks to leptons in the decays of $H$ and $A$ is
\begin{equation}
r_q^H = 3 \, \frac{m_b^2}{m_\tau^2} \tan^2 \alpha \tan^2 \beta
\label{eq:rqH}
\end{equation}
and
\begin{equation}
r_q^A = 3 \, \frac{m_b^2}{m_\tau^2}  \tan^4 \beta.
\label{eq:rqA}
\end{equation}
There is considerable uncertainty in the limit that the PAMELA
$\bar{p}$ data imposes on $r_q$.  For example, a limit of $r_q <
0.1$ can be satisfied by taking $\sin \alpha$ and $\sin \beta$
both $\lesssim 0.25$.  The limits on the mixing angles become much
more stringent for the case of heavier $H$ and $A$, where decays
to gauge boson or $\bar{t}t$ are possible.  In this case, $\sin
\alpha$ and $\sin \beta$ has to be sufficiently small. Although the
required small mixing angles remain within acceptable value (the $\tau$
yukawa coupling does not become large), for simplicity and concreteness
we will not consider this case for the rest of this paper.

\section{Astrophysics Signals}\label{astro}

DM particles in our galaxy and neighboring galaxies can annihilate
or decay into Standard Model (SM) particles leading to production
of cosmic rays such as electrons and positrons, protons and
anti-protons, photons and neutrinos, which could be observed at
the earth. Therefore, an observation of these cosmic rays
consistent with DM annihilation or decay could serve as
\emph{indirect} detection of Dark Matter. However, a given
framework for DM trying to explain a signal from one set
of experiments must also respect bounds set from all other
experiments.

In this section, we study the implications of the above framework
for cosmic ray signals - positrons and photons in particular.  We
also make comments about implications for the cosmic neutrino
flux from the Galactic Center (GC) at the end. In order to carry
out the analysis, we make the assumption, motivated in the
previous sections, that DM particles dominantly annihilate or
decay into higgs particles which have dominant $H_l$ components.
This can be guaranteed by postulating a symmetry (either discrete
or continuous)\footnote{Please see section \ref{explicit} for some
explicit models.} and mass range for $H$ and $A$ given in the
previous section.

We see that the robust and distinctive feature of the above
framework is production of tau leptons which subsequently give
rise to cosmic ray electrons and positrons, as well as photons and
neutrinos. However, the precise signal for cosmic rays depends on
model-dependent details. These can be broadly classified into five
classes. For annihilating DM, be it bosonic or fermionic, one has
the following: \ba\label{annpos} \chi\chi \rightarrow
HA\,(HH,\,AA) \rightarrow \bar{\tau}\tau\bar{\tau}\tau\ea For
bosonic annihilating DM, it is also possible to have:
\ba\label{annbpos} \chi\chi \rightarrow H^+ H^- \rightarrow
\bar{\tau}\tau\bar{\nu_{\tau}}\nu_{\tau}\ea whihc is forbidden at $s$-wave for
a majorana fermion DM by CP conservation. On the other hand, for
decaying DM, the signals for fermionic and bosonic DM are
different. For bosonic DM, there are two possibilities:
\ba\label{decaybposs} \chi &\rightarrow& HA (HH,\,AA) \rightarrow
\bar{\tau}\tau\bar{\tau}\tau\nonumber\\ \chi &\rightarrow&
\bar{\tau}\tau \ea The second possibility in (\ref{decaybposs})
can arise if the DM particle mixes with $H$ or $A$ which decays to
$\bar{\tau}\tau$. Finally, for fermionic decaying DM, one has the
possibilities: \ba\label{decayfposs} \chi &\rightarrow&
H\,(A)\,\nu \rightarrow \bar{\tau}\tau\nu\nonumber\\ \chi
&\rightarrow& H^{\mp}l^{\pm}\rightarrow
\tau^{\mp}\l^{\pm}\nu_{\tau};\;\;l\equiv e,\mu,\tau\ea

In section \ref{explicit}, we will construct some simple models
which exhibit all the above possibilities. Although a wide variety
of signals for cosmic rays can arise within this framework, it is
important to note that the annihilation and decay modes can be
related. More precisely, for the same given state (for example,
the 4 $\tau$ state in (\ref{annpos}) and (\ref{decaybposs})), the
signal for positrons and electrons in the annihilation mode for a
DM particle with mass $m_{\chi}$ and cross-section
$\langle\sigma\,v\rangle$ corresponds to that for a decaying DM
particle with mass $2\,m_{\chi}$ and lifetime $\tau_{\chi}$ given
by:\ba \tau_{\chi} \approx
\frac{m_{\chi}}{\rho_s\,\langle\sigma\,v\rangle} \ea where
$\rho_s$ is the dimensionful constant appearing in the DM
profiles\footnote{For a given profile, $\rho_s$ is constrained by
requiring that $\rho_{\chi}(r=8.5\,{\rm kpc})\approx 0.3\,{\rm
GeV/cm^3}$.}. The above holds true to a very good approximation
since the electrons and positrons observed at the earth come from
a short distance in the galaxy\footnote{more on this in section
\ref{positrons}.} where differences in the various DM profiles are
not important. Thus, the different dependence on the DM density
profile ($\sim \rho$ for decays versus $\sim \rho^2$ for
annihilations) does not have a big effect.

\subsection{Positrons ($\&$ Electrons)}\label{positrons}

In this subsection, we will estimate the positron fraction (to be
compared with PAMELA) and the total flux of electrons and
positrons (to be compared with ATIC) as a function of the mass of
the DM $m_{\chi}$ and that of the higgs particles $m_H$ and $m_A$,
consistent with the assumptions above. For concreteness, we will
show the results for the annihilation channel (\ref{annpos}) in
which DM annihilates to 4 $\tau$'s via two intermediate higgs
particles $H,A$. This can be easily translated to results for the
first decay mode in (\ref{decaybposs}) from arguments mentioned
above. Similarly, results for the second decay mode in
(\ref{decaybposs}) can be translated from that obtained for the
direct annihilation mode $\chi\chi \rightarrow \bar{\tau}\tau$ in
\cite{Cirelli:2008pk}. We will review this result and also comment
on the annihilation mode in (\ref{annbpos}) and the decay modes in
(\ref{decayfposs}) at the end of the subsection.

The cosmic-ray background of nuclei and electrons is believed to
originate from supernovae remnants but is not fully understood.
The nuclei and electron spectra is assumed to arise from an
injected flux which follows a power law as a function of energy,
and is then propagated through the galaxy within some
``propagation models". In the course of propagation through the
galactic medium, a secondary component of electrons and positrons
is generated by spallation of the cosmic rays on the interstellar
medium. The parameters of the source spectra and the propagation
models are constrained by fitting to astrophysical data. For
example, the nuclei source spectra and propagation parameters are
constrained by fitting to the proton data, the Boron-to-Carbon
(B/C) ratio and so on. Since background positrons are dominantly
generated from spallation of nuclei, this constrains the
background positron flux as well. The electron source spectra is
mostly constrained from experiments measuring the total electron
flux. Thus, in the absence of a complete theoretical understanding
of the processes involving the production and propagation of these
cosmic rays, there is a considerable amount of uncertainty in the
background electron and positron flux arising both from
uncertainties in the nuclei and electron source spectra,
production cross-sections, energy losses, as well as those from
parameters in the various propagation models. It turns out that
the uncertainty in the background electron spectrum is larger than
that in the propagation parameters at present. It is important to
keep these facts in mind when one tries to explain the data
observed by PAMELA and ATIC.

In addition to uncertainties in the background flux, there also
exist uncertainties in the ``signal" component assumed to arise
from the annihilation or decay of DM particles. Once the injection
spectrum of positrons at the source is specified, the primary
positron flux $\Phi^{prim}_{e^+}$ at the solar system arising from
DM annihilation in the Milky Way galactic halo is found by solving
a diffusion equation (with cylindrical boundary conditions for a
cylinder of half-height $L$=1-15 kpc and radius $R$=20 kpc) with a
source function given by:
\ba\label{source}
Q^{e^+}_{annih}(E',\vec{r'})&=&\frac{\rho_{\chi}^2(\vec{r'})}{2m_{\chi}^2}\,\langle
\sigma\,v\rangle\,\frac{dN_{e^+}}{dE'} (E')\nonumber\\
Q^{e^+}_{decay}(E',\vec{r'})&=&\frac{\rho_{\chi}(\vec{r'})}{m_{\chi}}\,\Gamma_{\chi}\frac{dN_{e^+}}{dE'}(E')
\ea
where $\frac{dN_{e^+}}{dE'} (E')$ is the injection spectrum of
positrons produced from DM annihilations or decay, and
$\rho_{\chi}(\vec{r})$ is the density profile of DM in our Galaxy.
As for the background, uncertainties exist in the propagation
parameters. Some of the most important parameters include the
half-height of the diffusion cylinder $L$, the parameters
characterizing the diffusion process - $K=K_0\,\beta{\cal
R}^{\delta}$, where $K_0$ is the diffusion constant, $\beta$ is
the velocity of the particle and $\mathcal{R}$ is its rigidity,
defined as $\mathcal{R}=|\vec{p}\,\mathrm{(GeV)}|/Z$ with $Z$ as
the atomic number, and the characteristic time for energy loss
$\tau_E$. Different sets of parameters $\{L,K_0,\delta\}$ exist
which are consistent with astronomical data such as the B/C ratio,
etc. The energy loss time $\tau_E$ has an uncertainty of about a
factor of 2 \cite{Olzem:2007zz}. The dependence on the DM profile
is weak since the positrons come from a short distance (${\cal
O}(1 {\rm kpc})$) where the different DM profiles are quite
similar. However, the average local DM density, i.e.
$\rho_{\chi}(r=8.5\,{\rm kpc})$ is itself uncertain by a factor of
2 \cite{Amsler:2008zzb}. From (\ref{source}), we see that this
uncertainty can be accommodated by a simple rescaling of the
cross-section (or decay width).

\begin{figure}[ht]
\centering
\resizebox{12cm}{!}{\includegraphics[0,0][507,335]{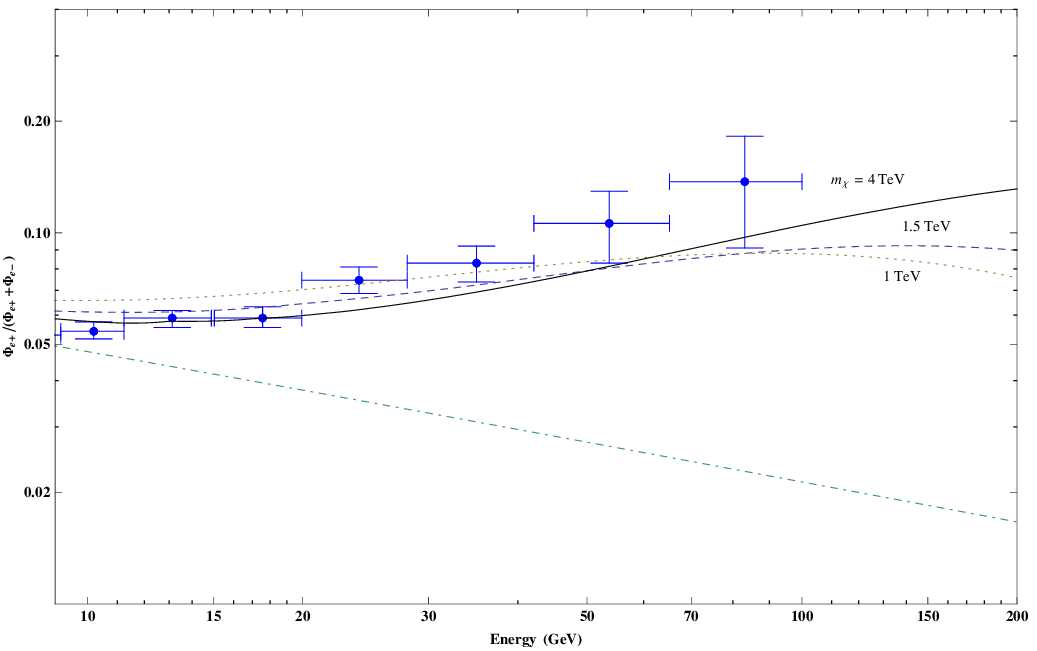}}\\
\resizebox{12cm}{!}{\includegraphics[0,0][446,286]{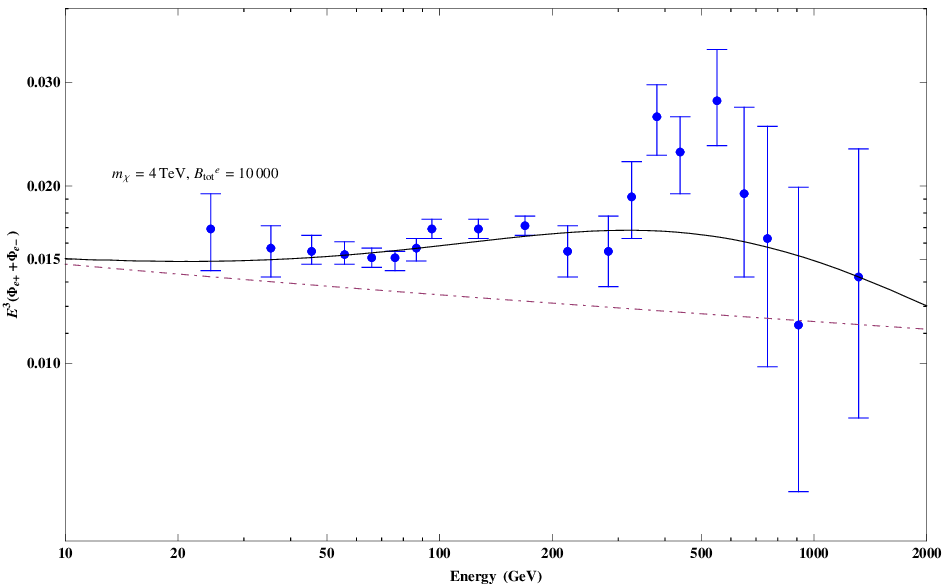}}
\caption{\footnotesize{Results for the positron fraction ({\bf
Top}) as a function of energy for DM masses $m_{\chi}=1$ TeV ({\bf
dotted}), $m_{\chi}=1.5$ TeV ({\bf dashed}) and $m_{\chi}=4$ TeV
({\bf black}) and $m_{H},m_A=100$ GeV, in the annihilation mode
(\ref{annpos}) with ``boost factors" $B^e_{tot}$ given by 1200,
1950 and 10000 respectively. The result for
$E^3(\Phi^{total}_{e^-}+\Phi^{total}_{e^+})$ ({\bf Bottom}) as a
function of energy is only shown for $m_{\chi}=4$ TeV. The
dot-dashed curve in both plots stands for the background. The
background electron spectral index $\alpha$ is taken to be 3.04
and  the MED propagation model \cite{Delahaye:2007fr} is used.
Also, the reference values of the local DM density $\rho_0$ and
$\tau_E$ are taken as 0.26 ${\rm GeV/cm^3}$ and $10^{16}$ seconds
respectively. The boost factor $B^e_{tot}$ is explained below
(\ref{boost}).}} \label{pamatic}
\end{figure}

Keeping the above facts about the signal and background fluxes in mind, we have estimated the positron fraction and the total flux of electrons and positrons. For concreteness, we have used the parameters for the MED propagation model for the background and signal fluxes, as defined in \cite{Delahaye:2007fr}. The background electron spectral index $\alpha$ defined by $\Phi^{bkg}_{e^-} \sim E^{-\alpha}$ is taken to be 3.04, which
is quite reasonable and is consistent with observations of all experiments \cite{Harnik:2008uu}. The normalization of the background electron spectrum is determined by a similar procedure as described in \cite{Harnik:2008uu}. Finally, we have used the Bessel approach to compute the Green's function of the diffusion equation. In particular, we have used an approximation to the Green's function found in \cite{Ibarra:2008jk} to solve the equation. The results of our analysis are shown in Figure
\ref{pamatic}.

We see from Figure \ref{pamatic} that the results for the positron
fraction and the total electron and positron flux are sensitive to
the mass of the DM $m_{\chi}$ as well as the ``Boost factor" for
electrons and positrons $B^e_{tot}$. The results are qualitatively
consistent with that of \cite{Cholis:2008wq} who have looked at a
similar annihilation mode, although from a very different
theoretical motivation and with much lighter masses of
intermediate scalars. One finds that a DM particle with mass
$m_{\chi} \sim$ TeV may explain the PAMELA data; however the
explanation of \emph{both} PAMELA and ATIC data requires that
$m_{\chi}$ is about 4 TeV and $B^e_{tot} \approx 10000$. Also, it
turns out that the above results have a mild dependence on the
masses of the intermediate higgs particles $m_H, m_A$ as long as
50 GeV $\lesssim m_H,m_A \lesssim$  few 100 GeV, which is the
expected range of masses for $H$ and $A$. Note that since the
results depend on astrophysical parameters such as the electron
spectral index $\alpha$ and the propagation model parameters as
explained above, it is possible to fit the data for different (but
comparable) values of $m_{\chi}$ and boost factors $B^e_{tot}$ by choosing a different
combination of these parameters. Also, if the ATIC data is ignored, the best fit values of
the parameters $m_{\chi}$ and $B^e_{tot}$ will be slightly different than when both data sets
are taken into account. Hence, the above results
should only be taken as an estimate. We have not tried to optimize
the fit (by a $\chi^2$ analysis) by the choice of astrophysical parameters.

The boost factor $B^e_{tot}$, which is a combination of various
factors, deserves some explanation. More precisely, the boost
factor is given by:

\ba \label{boost}
B^e_{tot}=B_{\sigma v}\cdot
B^e_{clump}\cdot B_{\rho_0} \cdot B_{\tau_E}
\ea

Here, $B^e_{\sigma v}$ is the enhancement factor in the
cross-section compared to the ``standard" one $\langle \sigma
v\rangle_{std} = 3\times 10^{-26} cm^3/s$. $B^e_{clump}$
corresponds to the enhancement in the positron (electron) signal
due to clumpiness in the DM halo. Strictly speaking, $B^e_{clump}$
is a function of energy \cite{Lavalle:1900wn}. Over the relevant
energy range of 1-1000 GeV, $B^e_{clump}$ of order few is
reasonable. $B_{\rho_0}$ is a possible enhancement due to a factor
of two uncertainty in the local average DM density itself, as
mentioned earlier. Note that a factor of two in $\rho_0$ appears
as a factor of four in $B_{\rho_0}$ because the flux goes as
$\rho_{0}^2$. Finally, the factor of two uncertainty in the energy
loss time $\tau_E$ mentioned above can be folded into a possible
enhancement because the flux is directly proportional to $\tau_E$.
Thus, $B^e_{tot}\approx 10000$ can arise in many ways. In
particular, it is perfectly compatible with $B_{\sigma v} \lesssim
1000$. In addition, the boost factor for neutrinos will in general
be \emph{different} than that for positrons (electrons). All these
facts will be important when we look at constraints from photons
and neutrinos in the following subsections as well as the section
on explicit models in section \ref{explicit}.

Moving on to the other annihilation and decay modes, the energy
spectra for $\tau$s and $\nu$s in the annihilation mode
(\ref{annbpos}) is expected to be roughly the same as in
(\ref{annpos}) because $m_{\tau}$ and $m_{\nu}$ are both
negligible compared to $m_{H^{\pm}}$. This suggests that it should
be possible to fit the data for the same $m_{\chi}$ as for
(\ref{annpos}) but with twice the boost factor. However, due to
$SU(2)$ invariance, we expect $\chi\chi\rightarrow AA,~HH$ which
gives rise to four $\tau$, to also contribute with the same cross
section. Taking the sum of these modes to be the total cross
section and normalizing it to the standard thermal cross section, we
need a factor of $4/3$ relative to the $B^e_{tot}$ required for only
the four $\tau$ case. This implies $B^e_{tot} \approx$ 13300 for $m_{\chi} \approx 4$ TeV.

As explained earlier, the results for the annihilation mode
(\ref{annpos}) can be translated to the first decay mode in
(\ref{decaybposs}). A DM mass of 4 TeV with $B^e_{tot}=10000$
corresponds to decaying DM with $m_{\chi}=8$ TeV and lifetime
$\tau_{\chi}$ given by \footnote{$\sqrt{B_{\rho_0}}$ arises due to
the fact that the flux goes as $\rho_0^2$ for annihilations but as
$\rho_0$ for decays.}:
\ba \tau_{\chi} \approx
\frac{m_{\chi}}{2\,\rho_s\,\langle\sigma\,v\rangle_{std}B^e_{tot}}(\sqrt{B^e_{clump}\cdot
B_{\rho_0}} \cdot B_{\tau_E}) \approx 3.2\times10^{25}\,{\rm
s}\,(\sqrt{B^e_{clump}\cdot B_{\rho_0}} \cdot B_{\tau_E}) \ea

To get the results for the second decay mode in
(\ref{decaybposs}), one needs to translate the results obtained
for the direct annihilation mode $\chi\chi \rightarrow \tau\tau$
in \cite{Cirelli:2008pk}. From \cite{Cirelli:2008pk}, one finds
that the $\chi\chi \rightarrow \tau\tau$ mode gives a good fit to
the PAMELA and ATIC data for $m_{\chi} \approx 2$ TeV and
$B^e_{tot} \approx 3000$. This corresponds, for the second decay
mode in (\ref{decaybposs}), to $m_{\chi}=4$ TeV and lifetime
$\tau_{\chi}$ given by:

\ba \tau_{\chi} \approx 5.3\times10^{25}\,{\rm
s}\,(\sqrt{B^e_{clump}\cdot B_{\rho_0}} \cdot B_{\tau_E}) \ea

Finally, we comment on the fermionic DM decay modes in
(\ref{decayfposs}). \cite{Ibarra:2008jk} studied the first decay
mode in (\ref{decayfposs}) and found that $m_{\chi}$ between 600
GeV and 1 TeV can explain the PAMELA data. It was pointed out in
\cite{Nardi:2008ix} that this decay mode can also provide a good
fit to both PAMELA and ATIC.  The second decay mode in
(\ref{decayfposs}) is qualitatively different, since the lepton
$l$ is harder than the $\tau$. For $l=e,\mu$ this provides a
contribution to the spectrum which is steeper than that coming
from the $\tau$, implying that it should be possible to fit the
data with a lighter $m_{\chi}$ and a smaller boost factor.

To summarize, therefore, even though the precise signal for
astrophysics depends on model-dependent details, the robust
characteristic of the framework is that an annihilating (decaying) DM
particle with mass of 4 (2-8) TeV and $B^e_{tot} =
O(10000-13000)$ ($\tau_{\chi} = 10^{25-26}\,{\rm s}$) can explain the
PAMELA and ATIC results. We now study the consequences for cosmic
gamma rays and neutrinos which provide non-trivial constraints on
the allowed parameter space as well as give rise to potential
signals for future experiments.

\subsection{Photons}\label{photons}

In general, any model of DM which produces a significant number of
electrons and positrons in the local region of our galaxy in the
energy range 10-1000 GeV (to explain the PAMELA and ATIC signals),
is expected to dominate the production of electrons and positrons
in the galactic center (GC) since the density of DM is expected to
be much larger there. This will in turn give rise to a large yield
of photons from inverse compton scattering (ICS) in the energy
range 1-1000 GeV. In addition, there could be other mechanisms of
photon production from DM annhilations or decays, such as
final-state radiation (FSR) of photons from charged particle
production, DM Bremsstrahlung, or from $\pi^0$s produced from
$\tau$ decays.

The total yield of photons is higher for annihilations than for decays since the flux
$\Phi \sim \rho_{\chi}^2$ for annihilations, while $\Phi \sim \rho_{\chi}$ for decays.
This can be seen from the general expresssion for the differential photon flux in a solid
angle region $\Delta\Omega$:
\ba\label{photon-flux} &&(\frac{d\Phi_{\gamma}}{dE_{\gamma}})_{annih}(\Delta\Omega,E_{\gamma})= \frac{1}{4\pi}\frac{\langle \sigma v\rangle}{2m_{\chi}^2}\,\sum_i\,b_i(\frac{dN_{\gamma}}{dE_{\gamma}})_i\,\bar{J}_{annih}\,\Delta\Omega\\
&&(\frac{d\Phi_{\gamma}}{dE_{\gamma}})_{decay}(\Delta\Omega,E_{\gamma})= \frac{1}{4\pi}\frac{1}
{m_{\chi}\,\tau_{\chi}}\,\sum_i\,b_i(\frac{dN_{\gamma}}{dE_{\gamma}})_i\,\bar{J}_{decay}\,\Delta\Omega\nonumber\\
&&\bar{J}_{annih}=\frac{1}{\Delta\Omega}\,\int_{\Delta\Omega}d\Omega
\int_{los}\rho_{\chi}^2(r(s))\,ds;\;\;\;
\bar{J}_{decay}=\frac{1}{\Delta\Omega}\,\int_{\Delta\Omega}d\Omega
\int_{los}\rho_{\chi}(r(s))\,ds\nonumber
\ea
Here, $\bar{J}_{annih}$ ($\bar{J}_{decay}$) corresponds to the
integrated squared (linear) DM density profile along the
line-of-sight, and $(\frac{dN_{\gamma}}{dE_{\gamma}})_i$ is the
photon spectrum coming from DM annihilations or decays for channel
$i$ with branching ratio $b_i$. Because of the different
parametric dependence on $\rho_{\chi}$,
$\left(\frac{\bar{J}_{decay}}{\rho_{solar}\,r_{solar}}\right) \ll
\left(\frac{\bar{J}_{annih}}{\rho_{solar}^2\,r_{solar}}\right)$
especially for ``steep" profiles like NFW, etc. This implies that
the decay mode gives rise to a much weaker signal compared to that
for the annihilation mode. On the other hand, the decay mode
satisfies the existing constraints from various observations of
gamma rays much more easily than the annihilation mode. Table
\ref{JbarGC} lists the values of $\bar{J}_{annih}$ and
$\bar{J}_{decay}$ (normalized such that they are dimensionless)
for the GC for two qualitatively different profiles - the NFW and
Isothermal profiles.
\begin{table}[h!]\label{JbarGC}
\begin{tabular}{|c|c|c|c|}
\hline
  & $\frac{\bar{J}_{annih}^2}{\rho_{solar}^2\,r_{solar}}$ & $\frac{\bar{J}_{decay}}{\rho_{solar}\,r_{solar}}$ \\ \hline
NFW   & $\approx 15\times10^{3}$
      &  $\approx 28.9$\\
Isothermal   & $\approx 13$ & $\approx 5.7$\\
\hline
\end{tabular}
\caption{$\bar{J}_{annih}$ and $\bar{J}_{decay}$ for the NFW and
Isothermal profiles of the GC in the Milky Way for standard
choices of astrophysical parameters, as in \cite{Bertone:2008xr,
Nardi:2008ix}.}
\end{table}

Since the decay modes give rise to a much weaker signal, we will
only show results for the annihilation mode, the one in
(\ref{annpos})  in particular. All three sources mentioned above
will contribute to the total photon yield for our framework in
general. For photon energies $E_{\gamma}\lesssim 100$ GeV, the
contribution from ICS turns out to be the most important
\cite{Cholis:2008wq}. For $E_{\gamma} \gtrsim 100$ GeV, the
contribution from $\pi^0$ decay takes over and dominates over the
ICS and FSR contributions. However, since the quantitative
predictions for $E_{\gamma} \lesssim 100$ GeV are subject to
various uncertainties in the ICS signal from astrophysics, we do
not attempt to analyze the above energy regime in this work.
Having said that, GLAST/FERMI is expected to be quite sensitive in
this energy range \cite{ucla-talk}. So, a strong signal by
GLAST/FERMI will provide very strong evidence for significant high
energy electron (positron) production in the GC.

\begin{figure}[h!]
    \begin{tabular}{c}
      \resizebox{12cm}{!}{\includegraphics[0,0][507,335]{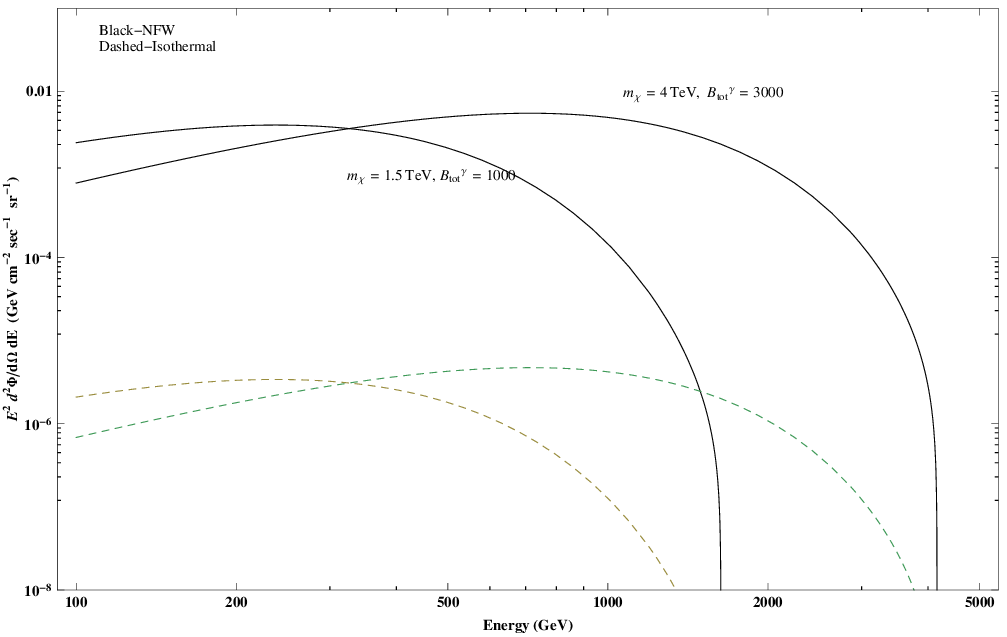}}\\
      \resizebox{12cm}{!}{\includegraphics[0,0][416,276]{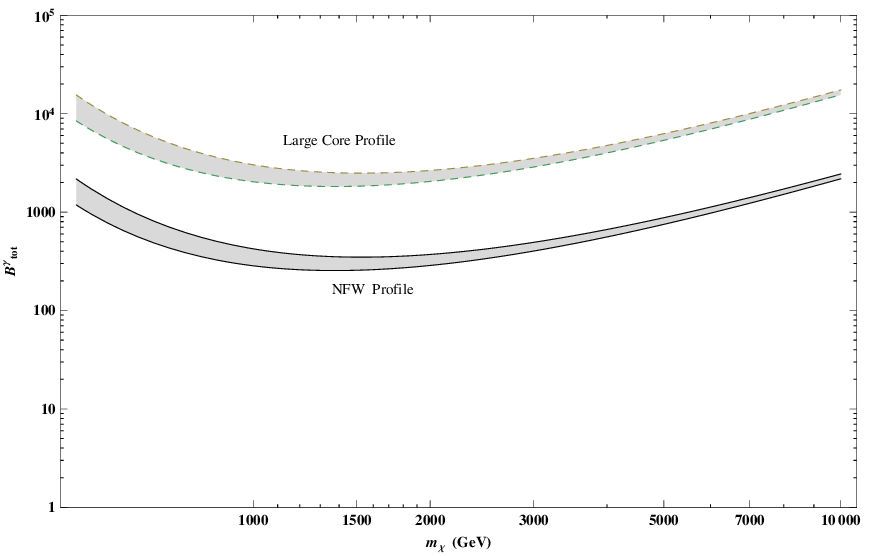}}
    \end{tabular}
\caption{\footnotesize{{\bf Top}: The photon intensity spectrum from the Galactic Center 
in the annihilation mode (\ref{annpos}) for DM masses $m_{\chi}=1.5$ TeV
($B^{\gamma}_{tot}=1000$) and $m_{\chi}=4$ TeV
($B^{\gamma}_{tot}=3000$) with $m_H,m_A=100$ GeV. The black curves stand for the
NFW profile while the dashed curves stand for the Isothermal
profile. The boost factor for photons $B^{\gamma}_{tot}$ is
explained below (\ref{boost-gamma}).} \\{\bf Bottom}:
\footnotesize{Upper bound on $B^{\gamma}_{tot}$, as computed in
(\ref{upperbound}), as a function of $m_{\chi}$ for the
annihilation mode (\ref{annpos}). The black curves stand for the
NFW profile while the dashed curves stand for the Large-Core
profile of Sagittarius. The parametrization in
\cite{Fornengo:2004kj} for $\frac{dN_{\gamma}}{dx}$ is used. The
shaded area corresponds to the uncertainty in extracting the photon spectrum
$\frac{dN_{\gamma}}{dx}$ from Monte-carlo simulations.}}
\label{future}
\end{figure}

For $E_{\gamma} \gtrsim 100$ GeV, one can study the detectability
of photon fluxes from DM annihilations or decays in Cerenkov
detector based experiments such as VERITAS 4 which is expected to
have a very high sensitivity in this energy range. The
differential flux sensitivity of VERITAS 4 is expected to be $\sim
8\cdot10^{-4}-\sim 2\cdot10^{-5}\,{\rm GeV cm^{-2} s^{-1}
sr^{-1}}$ in the energy range 100-1000 GeV \cite{ucla-talk}. The
top plot of Figure \ref{future} shows the predictions for the
photon intensity for the annihilation mode in the 100-1000 GeV
range, where $\pi^0$ decays dominate the photon yield. From the
figure, it can be seen that future experiments like VERITAS 4 have
a very good potential of detecting these very high energy gamma
rays in the annihilation mode, particularly for steeper profiles
like NFW. However, steep profiles may lead to some tension with
the flux observed by EGRET in the energy range 10-100 GeV. HESS
has also made observations of gamma rays coming from the GC
\cite{Aharonian:2006wh} and the Galactic Ridge (GR)
\cite{Aharonian:2006au}. However, these observations are not ideal
for DM observations because of the large contamination from gamma
ray point sources as well as from molecular gas which are not well known. In addition, the effective
$\bar{J}_{annih}$ relevant for these experiments is smaller because one has to subtract
``off-source" contributions \cite{Mardon:2009rc}. So, we do not attempt to analyze constraints from
these observations although one could presumably still place some conservative bounds.

Another set of important constraints for $E_{\gamma} \gtrsim 100$
GeV comes from HESS observations the dwarf spheroidal galaxy
(dSph) Sagittarius \cite{Aharonian:2007km} which is believed to
have negligible foregrounds, and hence is suitable for DM
observations. Observations of other nearby galaxies from MAGIC,
CANGAROO and WHIPPLE give similar constraints, so we will just
study the constraints from HESS.
HESS reports an upper bound on the integrated gamma ray flux $\Phi_{\gamma}^{max}= 3.6\times10^{-12}\,{\rm cm^{-2}\,s^{-1}}$ for $E_{\gamma} > 250$ GeV \cite{Aharonian:2007km}. From this, the following upper bounds on the DM annihilation cross-section (decay width) can be derived: \ba \label{upperbound}\langle\sigma\,v\rangle_{max} &=& (\frac{8\pi\Phi_{\gamma}^{max}\,m_{\chi}^2}{\bar{J}_{annih}\Delta\Omega\,\bar{N}_{\gamma}});\;\Gamma_{max} = (\frac{4\pi\Phi_{\gamma}^{max}\,m_{\chi}}{\bar{J}_{decay}\Delta\Omega\,\bar{N}_{\gamma}})\nonumber\\
{\rm where}\; \bar{N}_{\gamma} &=& (\int_{250\,{\rm
GeV}}^{m_{\chi}}\,\frac{dN_{\gamma}}{dE_{\gamma}}\,dE_{\gamma})\ea
where $\Delta \Omega = 2\times10^{-5}$ is the HESS solid angle
region. The quantity $\bar{N}_{\gamma}$ in (\ref{upperbound}),
{\it viz.} the number of photons above $\sim 250$ GeV, can be
estimated from the differential photon spectrum, as in
\cite{Fornengo:2004kj}. At present, there is considerable amount
of uncertainty in the DM density profile, which can vary from a
cusped profile (with various allowed values of the cusps) to cored
power-law profiles (with various values of the power-law
exponent). We will look at two qualitatively different profiles, a
large-core profile and an NFW profile, whose $\bar{J}$ (normalized
such that they are dimensionless) values are listed in Table
\ref{JbarSg}. As for the GC, one finds that
$\left(\frac{\bar{J}_{decay}}{\rho_{solar}\,r_{solar}}\right) \ll
\left(\frac{\bar{J}_{annih}}{\rho_{solar}^2\,r_{solar}}\right)$.
Therefore, we will only study the constraints for the annihilation
mode in (\ref{annpos}).
\begin{table}[h!]\label{JbarSg}
\begin{tabular}{|c|c|c|c|}
\hline
  & $\frac{\bar{J}_{annih}}{\rho_{solar}^2\,r_{solar}}$ & $\frac{\bar{J}_{decay}}{\rho_{solar}\,r_{solar}}$ \\ \hline
NFW   & $\approx 1\times10^3$ &  $\approx 1.3\times10^{-4} $\\
Large Core   & $\approx 1.4\times10^{2}$ & $\approx 1.2 \times 10^{-5} $\\
\hline
\end{tabular}
\caption{$\bar{J}_{annih}$ and $\bar{J}_{decay}$ for the NFW and
Large-core profiles of the Sagittarius dwarf spheroidal galaxy for
standard choices of astrophysical parameters, as in
\cite{Bertone:2008xr, Nardi:2008ix}.}
\end{table}

The bottom plot of Figure \ref{future} shows the upper bound on
$B^{\gamma}_{tot}$ as a function of $m_{\chi}$ for the NFW and
Large-Core profiles of Sagittarius in the annihilation mode
(\ref{annpos}). The area below the curves is consistent with the
HESS observations of Sagittarius. A correct interpretation of
these constraints requires some explanation. Taking into account
the possible enhancements in the signal from (\ref{upperbound}),
one finds that the boost factor for photons is given by: \ba
\label{boost-gamma} B^{\gamma}_{tot} = B_{\sigma v}\cdot
B^{\gamma}_{clump}\cdot
B_{\rho_0}=B^e_{tot}*(\frac{B^{\gamma}_{clump}}{B^{e}_{clump}\,B_{\tau_E}})\ea
where we have used (\ref{boost}). In general, the boost factors
arising from clumpiness in the halo is different for electrons
(positrons) and photons. This is because electrons and positrons
observed at the earth come from a short distance (since
they lose energy very quickly) whereas photons can come from very
far away. This implies that $B^e_{clump}$ is close to the
\emph{local} clumpiness boost factor $B^{local}_{clump}$ while
$B^{\gamma}_{clump}$ strongly depends on the direction of the
gamma-ray source relative to the earth \cite{Edsjo-talk}. So,
$B^{\gamma}_{clump}$ could be smaller than from $B^{e}_{clump}$ by
a factor around 1-10. Together with an uncertainty of a factor of
two in $B_{\tau_E}$, this could give rise to $B^{\gamma}_{tot}$
consistent with the bounds set by HESS. Note that the precise
value of $B^e_{tot}$ required to explain the PAMELA and ATIC data
can itself be changed by optimizing over the astrophysical
parameters consistent with the uncertainties. We have
utilized the discrepancy between $B^{\gamma}_{clump}$ and
$B^e_{clump}$ and the uncertainty in $B_{\tau_E}$ in the top plot
of Figure \ref{future}, and have taken $B^{\gamma}_{tot}$ as 1000
and 3000 for $m_{\chi}=1.5$ TeV and 4 TeV
respectively\footnote{These turn out to lie in between the upper
bounds set for the NFW and Large Core profiles, as seen in the
bottom plot of Figure \ref{future}.}.  On the other hand, if
$B^{\gamma}_{clump}$  is not sufficiently smaller than
$B^e_{clump}$, this would give rise to a tension between the
parameters required for explaining PAMELA and ATIC (in the
annihilation mode (\ref{annpos})) and the bounds set by the HESS
observations of Sagittarius. This tension is more severe for the
case of an NFW profile of Sagittarius (a little more than an order
of magnitude) .

To summarize, observations of gamma-rays from the Sagittarius
dwarf galaxy by HESS may provide strong constraints for the
annihilation mode (\ref{annpos}), primarily depending on the ratio
of clumpiness boost factors $B^{\gamma}_{clump}$ (in the
Sagittarius direction) and $B^e_{clump}$. On the
other hand, the decay modes in (\ref{decaybposs}) and
(\ref{decayfposs}) can easily satisfy these bounds. As is obvious,
the situation is reversed as far as prospects for future signals
are concerned. The annihilation modes (\ref{annpos}) and
(\ref{annbpos}) can give rise to observable signals from the GC,
which could be measured by VERITAS 4.

\subsection{Neutrinos}

A DM candidate which annihilates or decays to $\mu$'s or $\tau$'s
will also give rise to a significant flux of neutrinos and can
provide non-trivial constraints.  It was noted in
\cite{Hisano:2008ah, Liu:2008ci} that the neutrino flux from the
direction of the Galactic Center (GC) provides constraints for
both annihilation and decay modes, with constraints for
annihilation modes being much stronger (especially for steeper DM
profiles). The neutrinos coming from the direction of the GC can
be observed by detecting the muon flux induced by these neutrinos.
A detector in the northern hemisphere (such as Super-Kamiokande)
can detect upward going muons produced by neutrinos from the GC.
One has to take into account that the three flavors of neutrinos
oscillate into one another while traveling through the galaxy. As
was pointed out in \cite{Ritz:1987mh}, the observed
neutrino-induced muon flux is almost independent of $m_{\chi}$ for
annihilations (for fixed DM couplings, and small $m_{\chi}$). This is because both the
neutrino-nucleon cross-section and the muon range scale like
energy while the DM annihilation signal is proportional to
$1/m_{\chi}^2$. Thus, in contrast to the photon flux (see
(\ref{photon-flux})), the neutrino-induced muon flux is
proportional to the normalized second moment of the neutrino
energy spectrum, i.e. $\frac{d\Phi_{\mu}}{dE}\propto
(\frac{E}{m_{\chi}})^2\,\frac{dN_{\nu_i}}{dE}$. For decays, a
similar argument implies that the flux is proportional to
$m_{\chi}$. For heavier DM masses, the neutrino-nucleon
cross-section grows less steeply as well as the energy loss term
for muons starts becoming important, implying that the energetic
muon flux is relatively suppressed \cite{Mardon:2009rc}. It is
also important to note that the neutrino flux from the direction
of the GC is much less sensitive to the uncertainties in the DM
profile, especially if one looks at the GC over a large-size cone
centered at the GC. Then a large fraction of the total DM
annihilation signal is contained within the observed region.
Therefore, it is better to look at bounds set by Super-K for
large-size cones ($\sim 10^{\circ} - \sim 30^{\circ}$) around the
GC for robust results.

It was shown in \cite{Liu:2008ci} that the neutrino-induced muon
flux from direct DM annihilation to $\tau^+\tau^-$ for
$m_{\chi}\approx 2$ TeV, and $B^{\nu}_{tot}\approx 4500$ and an
NFW profile which provides a good fit to the PAMELA and ATIC data,
is slightly above the upper bound set by Super-K \cite{Desai:2004pq} for a cone-size
of about $10^{\circ}$ around the GC. When the DM annihilates via
higgs messengers to $\tau^+\tau^-$, the neutrinos are softer than
in the previous case. However, since the electrons (positrons) are
also softer, explaining the ATIC data requires that the DM mass in
this framework is quite heavy ($\approx 4$ TeV for the
annihilation mode) with a large boost factor $B^e_{tot}$ ($\approx
10000$). This effect tends to compensate the effect of soft
neutrinos mentioned above \cite{Mardon:2009rc} and one expects to get
approximately the same neutrino flux as for the direct
annihilation case\footnote{This has to be confirmed by explicit analysis.}. This naively implies that the neutrino-induced
muon flux for the annihilation mode (\ref{annpos}) with an NFW
profile of the GC is also above the bound set by Super-K. However,
as for gamma-rays, one has to keep in mind that $B^{\nu}_{tot}
\neq B^{e}_{tot}$ in general. The clumpiness boost factor
$B^{\nu}_{clump}$ is similar to that for photons because
neutrinos, like photons, hardly lose energy and come from far
away. Therefore, $B^{\nu}_{clump}$ is strongly direction-dependent
in general. To get more precise constraints from neutrinos coming
from the direction of the GC, one has to know the ratio of
$B^{\nu}_{clump}$ (in the direction of the GC)\footnote{It is
expected that $B^{\nu}_{clump}$ (GC direction)  $< B^{e}_{clump}
\approx B^{local}_{clump}$ \cite{Edsjo-talk} which would help in
relaxing the bounds.} and $B^e_{tot}$. As for photons, the results
for the annihilation mode (\ref{annpos}) for the isothermal
profile and for all the decay modes (for most profiles) are within
the bounds set by Super-K.

It is worth commenting about future experiments which are going to
look at the neutrino flux coming from various sources. The
Hyper-Kamiokande experiment, which is expected to have a
sensitivity bigger than Super-Kamiokande by about two orders of
magnitude, should be able to robustly detect a positive signal
from DM annihilations (and even decays in many cases) to neutrinos
through higgs messengers. For small cone-sizes ($\leq 2^{\circ}$),
future experiments like ANTARES \cite{Stolarczyk:2007ew} and
KM3neT \cite{Carr:2007zc} show exciting prospects for the
framework. For example, ANTARES and KM3neT should be able to
easily observe a significant neutrino signal from the GC for the
annihilation modes in (\ref{annpos}) and (\ref{annbpos}) for steep
DM profiles (like NFW) because $\bar{J}_{annih}$ increases rapidly
for small cone-sizes. The prospects for the decay modes are not as
promising since $\bar{J}_{decay}$ increases very slowly for small
cone-sizes. For high-energy neutrinos ($\gtrsim$ TeV) arising
within the framework, KM3neT may also be able to identify tau
neutrinos which would greatly help in suppressing the atmospheric
background since there are a lot fewer atmospheric tau neutrinos.

ICE-CUBE \cite{Ahrens:2003ix} does not look toward the GC, but
will instead look for a neutrino-induced muon flux arising from DM
annihilation or decays of DM particles which accrete in the earth
and in the sun. \cite{Delaunay:2008pc} has studied the prospects
for such a flux for direct DM annihilations to neutrinos and also
via annihilation to charged leptons such as taus. It was found
that direct DM annihilation to monochromatic neutrinos has good prospects
for ICE-CUBE. This implies that the production of neutrinos via
cascade decays through higgs messengers, as for modes
(\ref{annpos}), (\ref{annbpos}), (\ref{decaybposs}) and the second
mode in (\ref{decayfposs}), are also not promising. The first
decay mode in (\ref{decayfposs}), however, does lead to a monochromatic neutrino,
so one would expect that this provides much better prospects. As seen from Figure (2) in
\cite{Delaunay:2008pc}, discovery is possible for monochromatic
neutrinos if the enhancement factor in the
annihilation cross-section compared to $\langle \sigma v\rangle_{std}$ is 
$B_{\sigma v}\gtrsim$ 100-1000 because then the earth reaches equilibrium by the present time.   
However, the difference in the leptonic higgs framework is that the monochromatic neutrino signal arises from the \emph{decay} of DM particles rather than their annihilation. Since the denisty of DM particles inside the earth $\rho_{\chi}^{earth}$ is much larger than that in the galactic halo and since the flux in the decay mode scales as $\rho_{\chi}^{earth}$ in contrast to as $(\rho_{\chi}^{earth})^2$ for annihilations, the 
flux is greatly reduced \cite{Griest:1986yu} and hence does not provide good detection prospects at ICE-CUBE \footnote{This is true even if one assumes the "best-case" scenario that the earth has reached equilibrium at the present time due to a sufficiently large annihilation cross-section.}.

To summarize, observation of neutrinos originating from DM
annihilations (or decays) in the future will be crucial in greatly
strengthening the DM interpretation of the PAMELA and ATIC signals
over conventional astrophysical sources like pulsars since those
do not give rise to a large flux of neutrinos. The annihilation
modes (\ref{annpos}) and (\ref{annbpos}) provide stronger
constraints, but also provide a greater potential for
detectability in future experiments. Therefore, these deserve more detailed studies. It is
interesting to note that within the decay mode, neutrinos provide
a better opportunity for future detection of heavier DM compared
to photons since for a given DM density profile and decay width,
$\frac{d\Phi_{\mu}}{dE}\propto m_{\chi}$ for neutrinos which is not true for
the case of cosmic-ray photons.

\section{Higgs Physics - Collider Signals}\label{higgs-lhc}

We now move on to studying the Higgs sector in greater detail and
potential signals for the LHC. In order to do that, it is
important to study the higgs potential which is relevant for
understanding the production and decay modes of the various higgs
bosons. As mentioned earlier, we work within the framework of a
CP-invariant two-higgs doublet model. After electroweak symmetry
breaking, this gives rise to two CP-even higgs scalars $h$ and
$H$, a CP-odd higgs scalar $A$, and a charged higgs scalar
$H^{+}$. The couplings of these scalars to fermions was already
discussed in section \ref{leptonichiggs}. In the following
subsections, we first study the higgs potential within the
framework of a leptonic higgs and then discuss signals at the LHC.

\subsection{The Higgs Potential}\label{higgspot}

As stated in section \ref{leptonichiggs}, we consider a two higgs
doublet model in which a symmetry forces one of the higgs
doublets, $H_l$, to couple only to leptons, and the other higgs
doublet, $H_q$, to couple only to quarks. A simple example of such
a symmetry is a discrete $Z_2$ parity, $P_l$, under which $H_l$ is
odd, $H_q$ is even, the left-handed leptons are odd, while the
left-handed quarks and right-handed quarks and leptons are even.
Such an assignment enforces the couplings mentioned above.

The most general CP-invariant two higgs doublet potential consistent with the above parity
can be written as:
\begin{eqnarray}\label{eq:potential}
    V&=&-\mu_q^2(H^{\dagger}_qH_q)+\mu_l^2(H^{\dagger}_lH_l)+\frac{1}{2}\lambda_1(H_q^{\dagger}H_q)^2
         +\frac{1}{2}\lambda_2(H_l^{\dagger}H_l)^2+
  \lambda_3(H_l^{\dagger}H_l)(H_q^{\dagger}H_q)+
    \lambda_4(H_l^{\dagger}H_q)(H_q^{\dagger}H_l) +\nonumber\\
    &&(\frac{1}{2}\lambda_5(H_l^{\dagger}H_q)^2+h.c.).
\end{eqnarray}
It is not fine tuned to have a vacuum where the vev of the $H_l$
is smaller than that of $H_q$ by, say, a factor of three. A larger
hierarchy of vevs can be naturally obtained as follows.  The above
potential has an asymmetric phase where $H_q$ acquires a vev $v =
174$ GeV while $H_l$ has no vev. This phase of the two Higgs
doublet potential was studied in \cite{Barbieri:2006dq} for the
Inert Higgs Doublet model, where it was found that this phase has
a parameter space of comparable size to the standard phase, and
depends essentially on the sign of $\mu_l^2$.  Suppose one now
introduces a small, soft, parity breaking interaction in the
potential
\begin{equation}
\Delta V = -(\mu^2 H_q^{\dagger}H_l+h.c.).
\label{eq:DeltaV}
\end{equation}
Inserting the vev of $H_q$ into this interaction generates a
linear term in $H_l$, and therefore a vev for $H_l$ proportional
to the small symmetry breaking parameter $\mu^2$, which can
naturally be taken as small as desired. This will then guarantee
that $\tan \beta$ and $\sin \alpha$ are suppressed, as required
from the arguments in section \ref{leptonichiggs}. It is
convenient to parameterize the small parameter $\mu^2$ as
$\mu^2\equiv 2\epsilon\,v^2$ for later use. It is also helpful to
list the number of independent parameters. Equations
(\ref{eq:potential}) and (\ref{eq:DeltaV}) have eight parameters.
However, electroweak symmetry breaking gives rise to one condition
among these parameters, reducing the number of independent
parameters to seven. These can be taken as
$\{\lambda_1,\lambda_2,\lambda_3,\lambda_4,\lambda_5,\epsilon,t_{\beta}\}$.
In the linear approximation for $\sin \alpha$ and $\tan \beta$
(valid since both are small), the physical higgs masses and the
higgs mixing angle, $\sin \alpha$, can be computed in terms of
these parameters as:
\ba \label{approx-soln} m_H^2&\approx&2\,(\epsilon\,t_{\beta}^{-1})\,v^2;\;m_h^2\approx2(\lambda_1)\,v^2;\\
m_A^2 &\approx& 2\,(\epsilon\,t_{\beta}^{-1}-\lambda_5)\,v^2;\;m_{H^{\pm}}^2\approx2(\epsilon\,t_{\beta}^{-1}-\frac{1}{2}(\lambda_4
+\lambda_5))\,v^2;\nonumber\\ s_{\alpha} &\approx& t_{\beta}\,\frac{(\lambda_3+\lambda_4+\lambda_5-\epsilon\,t_{\beta}^{-1})}{(\lambda_1-\epsilon\,t_{\beta}^{-1})}\, \nonumber\ea
where $s_{\alpha}\equiv\sin \alpha$ and $t_{\beta}\equiv\tan
\beta$, and $\alpha$ and $\beta$ lie in the range $-\pi/2 \leq \alpha \leq \pi/2;\, 0 \leq \beta \leq
\frac{\pi}{2}$. For $\lambda_{1,2..,5}={\cal O}(1)$, and
$\mu^2\equiv \epsilon\,v^2 \ll |\mu_l|^2\sim|\mu_q|^2\sim v^2$,
the parameter $\epsilon = {\cal O}(1)\,t_{\beta}$. From
(\ref{approx-soln}), this implies that under these conditions all
higgs masses are comparable to each other, up to factors of ${\cal
O}(1)$. Also, depending on whether $(\epsilon\,t_{\beta}^{-1})$ is
smaller or greater than $\lambda_1$, $m_H$ could be lighter or
heavier than $m_h$. As we will show below, various choices of these ${\cal O}(1)$ numbers,
consistent with the constraints from LEP as reviewed at the end of section \ref{leptonichiggs}, can lead to a rich phenomenology at the LHC.


\subsection{Potential Signals at the LHC}

The leptonic higgs has very interesting collider phenomenology,
with distinctive signal characteristics that hold irrespective of
the explicit DM model.  If the ATIC data is confirmed,
then in our scheme the mass of the dark matter is too large for it
to be made at the LHC. Rather the LHC signals are encoded in the
Higgs messengers, and cover a wide range of dark matter models.
Although $H_q$ and $H_l$ mix, the mixing is required to be small
from dark matter considerations, so that the characteristics of
the leptonic higgs is kept intact. In particular, the couplings of
the mass eigenstate higgs bosons are shown in
(\ref{eq:SAhcouplings}).  Since $\alpha$ and $\beta$ are both
necessarily small, the higgs associated with EWSB, $h$, has SM
couplings to quarks but couplings to leptons that can be
considerably larger or smaller than in the SM. In the leptonic
Higgs sector, $H,A$ and $H^+$ all have enhanced couplings to
leptons and suppressed couplings to quarks.  This pattern of
couplings is quite unlike that of the MSSM where the fundamental
distinction is between up and down/lepton couplings, rather than
between up/down and lepton couplings.  The Higgs sector is as rich
as in the MSSM, and hence here we are only able to provide a
limited survey of the interesting signals. We choose to highlight
the decays of the neutral higgs bosons to pairs of $\tau$ leptons,
as this is the most direct link between cosmic-ray and LHC
signals.  In particular, we consider each of the cases:   $Z^* \,
\rightarrow \, HA \, \rightarrow \, 4 \tau$; $H,\,h \rightarrow
2\,\tau$ and $h\rightarrow 2A/2H \rightarrow 4\,\tau$, in some detail and also briefly mention
other possible LHC higgs signals such as the 8$\tau$ signal and the charged higgs signal.



\subsubsection{The $Z^* \, \rightarrow \, HA  \, \rightarrow \, \bar{\tau} \tau \bar{\tau} \tau$ Signal}

\begin{figure}[h!]
\resizebox{12cm}{!}{\includegraphics[0,0][392,271]{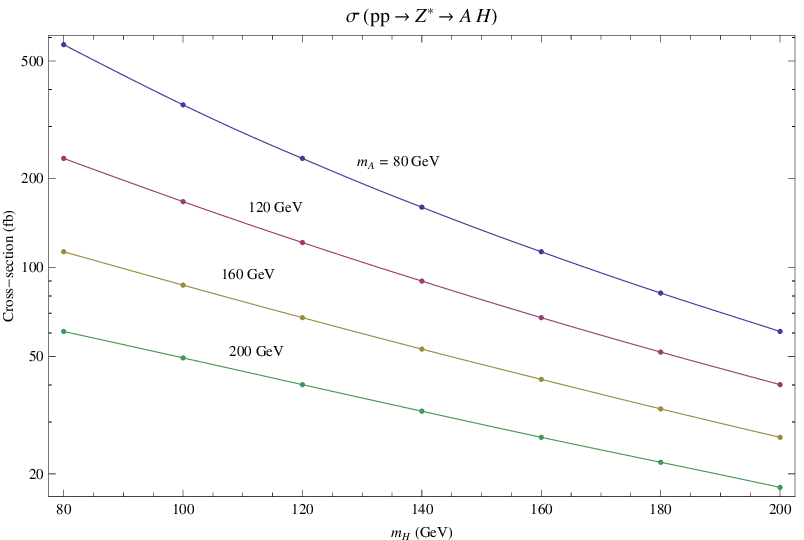}}
\caption{\footnotesize{Cross-section for $pp\rightarrow Z^*
\rightarrow H\,A\rightarrow 4\tau$ as a function of $m_H$ for
different values of $m_A$ at $\sqrt{s}=14$ TeV, computed with CalcHEP \cite{calchep}}.}
\label{cross-section}
\end{figure}

The LHC signal for $\bar{\tau} \tau \bar{\tau} \tau$ via $HA$
production is particularly robust because the $HA$ production
cross section is insensitive to the mixing angles $\alpha$ and
$\beta$, as they are both small, and $H,A$ decay to $\tau \tau$
with branching ratios larger than about 0.9.  The cross section
for $HA$ production at the LHC is dominated by
$s$-channel $Z$ exchange (Drell-Yan production)\cite{Dawson:1998py}. This is shown in Figure
\ref{cross-section}.

For example, with $m_H = 100$ GeV and $m_A = 80$ GeV the cross section is about
350 fb, while with $m_H = m_A = 160$ GeV, the cross section is
about 50 fb. This is the
same as the cross section for $HA$ production in the decoupling limit of the MSSM.
However, in the MSSM the $\bar{\tau} \tau$ branching ratio of $H$ and $A$ is never
larger than $0.08$, and is frequently significantly smaller,
depleting the $\bar{\tau} \tau \bar{\tau} \tau$ signal by at least
two orders of magnitude, whereas for the examples of $H$ and $A$
masses given above we expect about 10500 and 1500 $\bar{\tau} \tau
\bar{\tau} \tau$ events (before imposing any cuts) with an integrated luminosity of 30
fb$^{-1}$. The main background for this signal is expected to be diboson ($Z$) production
in which the $Z$'s decay to $\bar{\tau} \tau$. Given that the signal is independent of the mixing
angles $\alpha$ and $\beta$, and depends only on $m_H$ and $m_A$,
which we expect to be bounded by about $2 M_W$ and $m_H+m_Z$
respectively from earlier cosiderations, it will be very important to study a realistic
simulation of this signal with selection cuts and detector
effects. 

Before moving on, it is worth pointing out that the same Drell-Yan process will also lead
to $H^+H^-$ production with roughly the same cross-section; hence it could give rise to an observable signal
with sufficient luminosity for the dominant $H^+\rightarrow \tau^+\nu_{\tau}$ and $H^-\rightarrow \tau^-\nu_{\tau}$
channels. This suggests a completely different strategy to search for charged higgs bosons in contrast to that for the MSSM,
which focusses on production of charged higgs boson production in association with top quarks and then studying their
hadronic ($t\,\bar{b}$) and leptonic ($\tau^+\nu$) decay signatures.

\subsubsection{The $h,H  \, \rightarrow \, \bar{\tau} \tau$ Signal}

The two $\tau$ signal from higgs decays has been studied in the
SM. Although gluon fusion is the dominant production channel for
the higgs, the search strategy for the $h_{SM}\rightarrow
\bar{\tau}\tau$ mode consists of exploiting the vector boson fusion
(VBF) channel for $h_{SM}$ production, which has some distinct
features that help to suppress the otherwise large backgrounds.
Higgs production in this channel is usually accompanied by two
jets in the forward region originating from the initial quarks
from which vector bosons are emitted. Another feature is that no
color is exchanged in the central hard process, leading to low jet
activity in the central region. This is in contrast to most
background processes. So, jet tagging in the forward region
together with a veto of jet activity in the central region can
help in achieving a high signal significance.

We estimate the $\sigma_{VBF}\times BR(\bar{\tau}\tau)$ for $h$
and $H$, and compare it with the SM case for the same higgs mass.
Since $A$ does not couple to $WW$, one only has to consider $h$
and $H$. One finds:
\begin{eqnarray}\label{2tau}
\sigma_{VBF}(h) \times BR (h\rightarrow \bar{\tau}\tau) &\approx& [\sigma_{VBF}^{SM}\times BR(h_{SM}\rightarrow \bar{\tau}\tau)]\,
\frac{\left(\frac{\sin^2 \alpha}{\sin^2 \beta}\right)}{[1+\left(\frac{\sin^2 \alpha}{\sin^2 \beta}-1\right)BR(h_{SM}\rightarrow \bar{\tau}\tau)]}\\
\sigma_{VBF}(H) \times BR (H\rightarrow \bar{\tau}\tau) &\approx&
[\sigma_{VBF}^{SM}\times BR(h_{SM}\rightarrow \bar{\tau}\tau)]\,
\frac{\sin^2(\alpha-\beta)}{BR(h_{SM}\rightarrow \bar{\tau}\tau)}.
\nonumber
\end{eqnarray}
For $m_H, m_h \lesssim 150$ GeV, the existing studies for the SM
can be used to estimate the discovery potential in the proposed
framework \cite{cms-tdr}. In these studies, the mode in which one
of the $\tau$'s decays leptonically while the other decays
hadronically is analyzed in detail for an integrated luminosity of
$30\,fb^{-1}$. For this mode, $[\sigma_{VBF}^{SM}\times
BR(h_{SM}\rightarrow \bar{\tau}\tau \rightarrow lj)]$, with $l =
e/\mu$ and $j = \mbox{jet}$, ranges between 45 and 155 $fb$ for
$145 > m_{h_{SM}} > 115$ GeV. After imposing various selection
cuts to reduce the background, one gets between $\sim 4$ to $\sim
10$ events for the signal, compared to about $\sim 1.5$ to $\sim
3.5$ events for the background \cite{cms-tdr}. This gives rise to
a signal significance of about $3\sigma-4\sigma$ for 30 $fb^{-1}$
in the VBF production channel.

\begin{figure}[h!]
    \begin{tabular}{cc}
     \resizebox{8cm}{!}{\includegraphics[0,0][247,242]{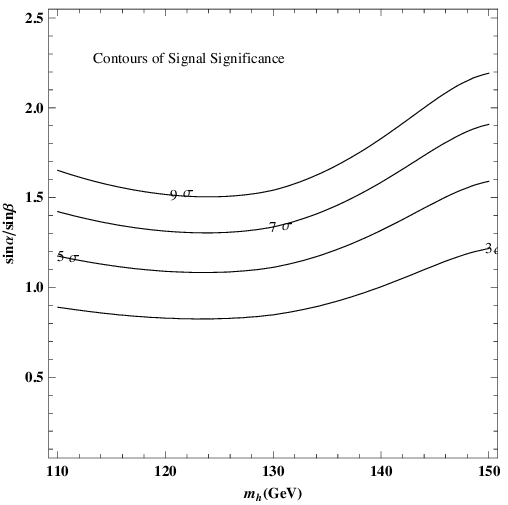}}&
     \resizebox{8cm}{!}{\includegraphics[0,0][247,242]{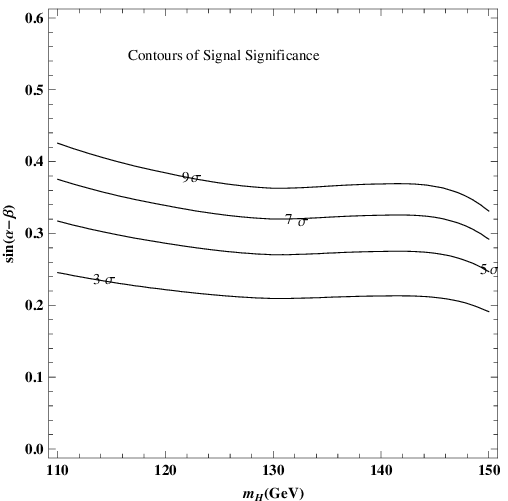}}\\
    \end{tabular}
\caption{\footnotesize{Signal Significance contours for the
$h\rightarrow \bar{\tau}\tau$ channel ({\bf left}) and for the
$H\rightarrow \bar{\tau}\tau$ channel ({\bf right}) for an integrated luminosity of 
30 $fb^{-1}$. Results from
Figure 14 in \cite{Abdullin:2005yn} have been used}.}
\label{twotau}
\end{figure}

For $h$ and $H$, the signal is modified as in (\ref{2tau}). First,
we identify a parameter region that gives large signals for both
$h$ and $H$.  A large value of $\sin \alpha$ relative to $\sin
\beta$ is required, while both angles must be small to account for
the DM signals. As a benchmark, one could take $\sin \alpha
\approx \alpha =0.35$ and $\sin \beta \approx \beta=0.05$ giving
rise to $\frac{\sin \alpha}{\sin \beta}\approx 7$, and
$\sin(\alpha-\beta)\approx 0.3$. From (\ref{approx-soln}),  such a
value of $\frac{\sin \alpha}{\sin \beta}$ can be obtained by
choosing ${\cal O}(1)$ numbers for the $\lambda$'s and $\epsilon$.
For the same range of masses for the higgses $h,H$ as used for $h_{SM}$ in 
\cite{cms-tdr}, one finds for
the benchmark values: \ba \sigma_{VBF}(h) \times BR (h\rightarrow
\bar{\tau}\tau) &\approx& (10-20)\,[\sigma_{VBF}^{SM}\times
BR(h_{SM}\rightarrow \bar{\tau}\tau)]\nonumber\\\sigma_{VBF}(H)
\times BR (H\rightarrow \bar{\tau}\tau) &\approx&
(1.1-3)\,[\sigma_{VBF}^{SM}\times BR(h_{SM}\rightarrow
\bar{\tau}\tau)].\ea Thus, assuming that the results for the
number of signal events scale in a simple way, one gets a huge
signal significance of $\sim 40\,\sigma$ to $\sim 60\,\sigma$ for
the $h$ mode, and a significance of $\sim 4.4\sigma$ to $\sim
9\sigma$ for the $H$ mode, for $115 < m_{h}, m_{H} < 145 $ GeV
in the VBF production channel for a luminosity of 30 $fb^{-1}$.

Of course, the benchmark values have been chosen to maximize the
signal siginificance. To get a better idea of the allowed
parameter space, in Figure \ref{twotau} we show statistical
significance contours in the $m_h-(\sin\alpha/\sin\beta)$ plane
for $h$ and in the $m_H-\sin(\alpha - \beta)$ plane for $H$. 
As long as $|\alpha|>\beta$ \footnote{$\beta$ can
always be chosen to lie in the first quadrant.}, while both are
``small" ($|\alpha|\lesssim 0.3$), it seems reasonable to claim
that the $\bar{\tau}\tau \rightarrow lj$ channel (especially for
the $h$ mode) provides an extremely robust signal and can be
readily discovered at the LHC for higgs masses $\lesssim 150$ GeV.
The above analysis is valid for $m_h, m_H \lesssim 2\,m_W$. For
$m_h$ much greater than $2\,m_W$, the $\bar{\tau}\tau \rightarrow
lj$ channel does not seem promising as in this case $h$ decays
dominantly to $WW$ and $ZZ$, which become the best modes to search
for.

\subsubsection{The $h  \, \rightarrow \, HH, AA  \, \rightarrow \, \bar{\tau}\tau\bar{\tau}\tau$ Signal}

A 4$\tau$ signal is possible when $m_h > min\{2\,m_A,2\,m_H\}$,
with $h\rightarrow AA$ or $HH$ and subsequent decays into four
$\tau$. The prospects for discovery through this mode are better
when $m_h \lesssim 130$ GeV; otherwise the $t\bar{t}$ background
becomes quite large and the branching ratio of $h\rightarrow A$ or
$H$ goes down as well. So, we will first analyze the case
$130\,{\rm GeV}\gtrsim m_h > min\{2\,m_A,2\,m_H\}$. The four
$\tau$ final state is expected to have smaller background than the
four $b$ final state. Again, the most favored production channel
is VBF for the same reasons as in the previous subsection. The
higgs-strahlung production channel with leptonic gauge boson decay
can also provide a nice trigger and better handle on the
background. However, the cross section for this channel is much
lower than that for VBF.

A full simulation study of the 4$\tau$ channel with  the $\tau$'s
decaying into 4 $\mu + 4\nu_{\tau} + 4\nu_{\mu}$ is currently
under study at ATLAS \cite{:2008uu} in the context of an NMSSM
model in the VBF production channel of a SM-like higgs with the
higgs decaying to $AA$ followed by the decay to 4$\tau$'s, just as
in the proposed framework. It requires three leptons to be
observed and triggers on one or two high $p_T$ leptons. CMS, on
the other hand, is investigating the mode 4 $\tau \rightarrow
\mu^{\pm}\mu^{\pm}\tau^{\mp}_{jet}\tau^{\mp}_{jet}$ containing two
same sign muons and two $\tau$ jets in the context of the same
NMSSM model framework \cite{:2008uu}. Although a full study is
currently unavailable, some studies of benchmark points in the
NMSSM have been performed \cite{Assamagan:2004mu}. In these
studies, the 4$\tau$ channel is studied when two $\tau$'s decay
hadronically and the other two decay leptonically. For example, a
benchmark NMSSM model, which seems to be quite similar to that for
the proposed framework as far as the 4$\tau$ channel is concerned,
is given by: $\{m_{h_1}=120\,{\rm GeV},m_{A_1}=7\,{\rm
GeV},BR(h_1\rightarrow A_1A_1)=0.99,BR(A_1\rightarrow
\tau\tau)=0.94\}$ where the $h_1$ behaves as a SM-like higgs as
far as coupling to quarks and gauge bosons are concerned. A
preliminary study of this benchmark model indicates that a signal
significance of about 20 $\sigma$ could be obtained for an
integrated luminosity of $300\,fb^{-1}$ \cite{Assamagan:2004mu}.

In the proposed framework, one has a similar situation as the
above NMSSM benchmark model for the 4$\tau$ signal. The parameters
$m_h$, $BR(h\rightarrow AA)$ and $BR(A\rightarrow \tau\tau)$ are
very similar to their counterparts for the above NMSSM benchmark
model for $m_h < 2\,m_W$. One possible exception is the mass of
$A$ which could be much heavier than that for the NMSSM model
while still being consistent with a significant 4$\tau$ signal.
For $m_h=120$ GeV, one expects to get a similar statistical
significance ($\sim 20\,\sigma$) as in the above benchmark model
for 300 $fb^{-1}$. Since the proposed framework can have much
heavier $m_A$, the two $\tau$'s from the decay of $A$ are better
separated implying that one could presumably get a better
significance by utilizing this feature. If the signal and
background are assumed to scale in a simple way, this would imply
a statistical significance greater than 5 $\sigma$ even for a
luminosity of 30 $fb^{-1}$. Therefore, for $m_h \lesssim 130 $
GeV, the $4\tau$ channel may be used to make a discovery.

For heavier $h$, i.e. for $m_h \gtrsim 130$ GeV, the above search
strategy is not as promising because of the huge $t\bar{t}$
background, which begins to rise sharply at about
$M_{\tau\tau\tau\tau}=140$ GeV \cite{Assamagan:2004mu}. Further
studies from ATLAS and CMS are therefore needed to claim any
significance for $m_h \gtrsim 130$ GeV. Finally, it is worth
mentioning that in addition to studies at ATLAS and CMS, there is
a study based on the proposed forward proton detector at the LHC,
the so-called FP420 project \cite{Gunion}. This proposal utilizes
the diffractive production $pp\rightarrow pph$ and detects protons
in the final state. The claim from this study is that the final state basically
consists of events with no backgrounds, implying that the masses
of $h$ and $A$ can be determined on an event-by-event basis.

Finally, we would like to point out an interesting possibility arising from $h$ pair-production.
If $m_h > min\{2\,m_A,2\,m_H\}$ as above, $h$ pair-production could lead to a spectacular 8 $\tau$
signal at the LHC. Higgs pair-production in the SM is dominated by gluon fusion \cite{Gianotti:2002xx}.
For example, for a 120 GeV higgs ($h_{SM}$), the cross-section for $h_{SM}$ pair-production is about 35 fb. The cross-section for $h$ pair production in our framework is the same as that for the SM. This implies that one could
get about 1000 8$\tau$ signal events (before imposing any cuts) for 30 $fb^{-1}$. Again, a detailed
analysis of this channel would be quite interesting.

\section{Some simple DM Models with Leptonic Higgs}\label{explicit}

We now complete the picture in Figure \ref{cartoon} by explicitly constructing
the DM sector and the couplings of this sector through the leptonic Higgs. 
To demonstrate the idea, we study three models: the $L-L^c-N$ model
where the DM particle is Majorana fermion, and models where the DM particle is a scalar, either 
singlet or electroweak doublet. In each of these models,
we consider both annihilation and decay modes.

\subsection{$L-L^c-N$ DM Model}

The dark sector in this model consists of a
vector-like pair of lepton doublets $L$, $L^c$ and a sterile
neutrino $N$. $L$ has exactly the gauge charges of the lepton doublet
in the SM. The model has two $Z_2$ parities - a chiral lepton
parity $P_l$ introduced in section \ref{higgspot} under which
$H_l$ and all the lepton doublets $L$, $L^c$ and $L_i$ are odd while
all other fields are even, and a dark parity $P_D$ under which
only particles in the dark sector, $L$, $L^c$ and $N$, are odd and
all other particles are even. $P_D$ guarantees that the lightest
particle among $L$, $L^c$ and $N$ is stable and can be a dark
matter candidate if it is neutral.

This model is similar to the model proposed in \cite{Enberg}
except that it has an additional higgs which couples predominantly to leptons.
The gauge couplings of $L$ and $L^c$ are
standard, and the other renormalizable interactions involving
the dark sector are:
\begin{eqnarray}\label{eq:LLN}
\Delta{\cal L}&=&\eta_1 H_l^{\dagger}LN+\eta_2
    H_l^{T}L^cN +m_L LL^c + \frac{1}{2}m_N\,N^2
\end{eqnarray}
The nature of the DM candidate depends on the spectrum of the dark
sector which in turn is determined by the mass parameters $m_L$, $m_N$ and
the higgs vev $\langle H_l\rangle =v_l$. The charged components $\chi^{\pm}$ in
$L$ and $L^c$ form a Dirac fermion with mass $m_L$. $N$ and the
neutral components of $L$ and $L^c$ mix after electroweak symmetry
breaking. The three neutral Majorana mass eigenstates
$(\chi_1,\chi_2,\chi_3)$ and their corresponding masses
$(m_1,m_2,m_3)$, are given by:

\begin{eqnarray}
\left(%
\begin{array}{c}
  \chi_3 \\
  \chi_2 \\
  \chi_1 \\
\end{array}%
 \right)&\sim &\left(%
\begin{array}{ccc}
   1& \frac{(\eta_1-\eta_2)v_l}{\sqrt{2}(m_N+m_L)}&\frac{(\eta_1+\eta_2)v_l}{\sqrt{2}(m_N-m_L)}\\
   O(v_l/m_N)&1 &\frac{(\eta_1^2 - \eta_2^2)v_l^2}{4 m_L (m_N + m_L)} \\
   O(v_l/m_N)& -\frac{(\eta_1^2 - \eta_2^2)v_l^2}{4 m_L (m_N -
   m_L)}&1\\
\end{array}%
\right)\left(%
\begin{array}{c}
  N \\
  (L-L^c)/\sqrt{2} \\
  (L+L^c)/\sqrt{2} \\
\end{array}%
\right)
\end{eqnarray}
\begin{eqnarray}\label{eq:spect}
    m_3&\sim &m_N+O(v_l^2/m_N) \nonumber\\
      m_2&\sim &m_L+\frac{(\eta_1-\eta_2)^2}{2(m_N+m_L)}v_l^2\nonumber \\
     m_1&\sim &m_L-\frac{(\eta_1+\eta_2)^2}{2(m_N-m_L)}v_l^2 \nonumber\\
\end{eqnarray}

If $m_N > m_L$, the spectrum becomes:
\begin{equation}
    m_{\chi_1}<m_{\chi^{\pm}}<m_{\chi_2}<m_{\chi_3}
\end{equation}
and one gets doublet dark matter. The splitting due to electroweak
symmetry breaking guarantees that the lightest $P_D$ odd particle is
always neutral, and the splitting between $\chi_1$ and $\chi_2$ allows
the model to evade the bound from direct detection by suppressing elastic scattering
through coupling to a $Z$ boson\footnote{Elastic scattering through the higgs is also suppressed due to the suppressed couplings of the leptonic higgs to quarks.}.

On the other hand, if $m_N < m_L$ one gets:
\begin{equation}
    m_{\chi_3}<m_{\chi_1}<m_{\chi^{\pm}}<m_{\chi_2}
\end{equation}
and one gets singlet dark matter. Therefore, this model allows the DM particle
to be either a singlet or electroweak doublet. It is also possible to have
decaying dark matter in this model if the dark parity $P_D$ is broken by a small amount. We now
discuss both DM annihilations and decays within this model. For simplicity, we assume
that only one of the modes is responsible for the cosmic-ray signals although in principle
it is possible that both modes are comparable.

\subsubsection{Annihilating Dark Matter}

The phenomenological consequences in the annihilation mode for singlet and doublet DM
are different. In addition to the differences in the cross-section, a doublet DM particle has an unsuppressed coupling to the $Z$ boson in contrast to a singlet one which doesn't couple to the
$Z$. For singlet DM, it turns out that the dominant annihilation channel in the $s$-wave
is $\chi\chi\rightarrow H\,A$ by a $t$-channel exchange of the heavier neutral
partner of $L$ and $L^c$. All other $s$-wave channels, such as $ZZ$,
$WW$, $Z$-higgs, $W$-higgs, and fermion pairs ($f\bar{f}$) are
suppressed because of the tiny mixing between the doublet and
singlet components. The $H^+ H^-$, $HH$ and $AA$ channels are also
suppressed due to CP invariance. For doublet DM however,
in addition to the $\chi\chi\rightarrow H\,A$ channel as before, annihilation to $ZZ$ is also potentially relevant because of the unsuppressed coupling mentioned above.
Since $Z$ has a large hadronic branching ratio, the branching ratio of the doublet DM
annihilating to $ZZ$ has to be much smaller than that to $H\,A$ in order for it to be
phenomenologically viable. The annihilation cross-sections for doublet DM
are given by: \ba \label{crosssection}
\sigma\,v^{(D)}\,(\chi\chi\rightarrow HA)_{s}
&=&\frac{(\eta_2^2-\eta_1^2)^2}{16\pi\,m_{N}^2(1+\frac{m_L^2}{m_N^2})^2}(1+O(\frac{m_{H,A}^2}{2m_{L}^2}))\nonumber\\
\sigma\,v^{(D)}\,(\chi\chi\rightarrow ZZ)_{s} &=&
\frac{g^4}{32\pi\cos^4{\theta_W}\,m_{L}^2}\,(1+O(\frac{m_Z^2}{m_{L}^2}))\nonumber
\ea

From the argument above, this implies a
lower bound on the coupling $(\eta_2^2-\eta_1^2)$ (for a given
$m_{\chi}$ and $m_{\chi'}$). We know from section \ref{astro} that
in order to fit the PAMELA and ATIC
data, the DM mass $m_{\chi}$ is required to be around 4 TeV and the
total boost factor for positrons (electrons) $B^e_{tot}$ needs to be around $10^4$.
Taking into account the local fluctuation of the DM density and other
uncertainties which could naturally give rise to a boost factor of about 10,
this would imply an enhancement of about $10^3$ from the cross-section
itself. For the doublet case, this naively implies that $\eta_2^2-\eta_1^2 \sim
40$. However, it turns out that there is a mild Sommerfeld enhancement in this model
due to formation of a wimponium bound state by $W$-exchange. This enhances the cross-section
by a factor of about 10 if the mass splitting between $\chi_1$ and $\chi^+$ is
less than $0.1$ GeV \cite{Hisano}. From eq.(\ref{eq:spect}), this
small mass splitting can be achieved with $v_l < 5$ GeV
corresponding to $\sin\beta \sim 3\times
10^{-2}$ ($y_{\tau}\sim 0.3$). Therefore, a perturbative cross-section (\ref{crosssection})
which is smaller than that naively required (without the Sommerfeld enhancement)
by a factor $\sim 10$ is also allowed. This in turn
implies that a much smaller $\eta_2^2-\eta_1^2$ is needed. For example, $\eta_1 << \eta_2 \sim
3.5$ can explain the PAMELA and ATIC data.

Moving on to the singlet case, the annihilation cross-section to $H\,A$ is
given by: \ba \label{crosssection2}
\sigma\,v^{(S)}\,(\chi\chi\rightarrow H\,A)_{s}
=\frac{(m_{H}^2-m_{A}^2)^2}{4m_{N}^2m_{L}^2}
\sigma\,v^{(D)}\,(\chi\chi\rightarrow H\,A)_{s}\nonumber \ea
This is suppressed compared to that for the doublet case by typically a factor
of $10^{-4}$, which would require the yukawa couplings $\eta_1,\eta_2$ to be much larger than
the strong coupling limit to explain the observed data. Therefore, we do not discuss this case.

\subsubsection{Decaying Dark Matter}

It is interesting to consider this model in the decay mode since the annihilation mode
is subject to more stringent constraints from cosmic-ray photons and neutrinos as explained in section \ref{astro}. In order to have decaying
DM in this model, the following $P_D$ breaking
terms can be added to the Lagrangian ({\ref{eq:LLN}}) consistent with all gauge symmetries:
\begin{eqnarray}\label{eq:LLNdecay}
\Delta{\cal L}_{decay}&=& \delta_{1i} H_l^{\dagger}L_iN +
\delta_{2i}
    H_l^{T}Le_i^c+\delta_{3i}H_l^{\dagger}L^ce_i^c+\delta'_{i}m_L L_iL^c+h.c.
\end{eqnarray}
$\delta$'s are required to be extremely small to allow the DM lifetime to be around $10^{25-26}$ seconds. The particular form
of the above interaction and the smallness of $\delta$'s can be explained by
symmetry arguments in several different ways.

The mass mixing term in (\ref{eq:LLNdecay}) with coefficient $\delta'_i$
dominates the decay process. Therefore, the parity $P_{l}$ has to be suitably extended to give rise to sufficiently long decay lifetimes. One simple possibility is to extend $P_l$ to the full chiral leptonic parity
$P_{l_l}\times P_{l_r}$ where $H_l$ is odd under both, $L_i$ and
$N$ are odd under only $P_{l_l}$, while $e_i^c$, $L$ and $L^c$ are
odd under only $P_{l_r}$. Then, the leading non-renormalizable operators are:
\begin{equation}
    \frac{\phi_l\phi_r\phi_D}{M_{UV}^3}(c_{1i} H_l^{\dagger}L_iN +
    c_{2i}H_l^{T}Le_i^c+c_{3i}H_l^{\dagger}L^ce_i^c)+c_{4i}\frac{\phi_l\phi_r\phi_D}{M_{UV}^2} L_iL^c
\end{equation} where $\phi_l$, $\phi_r$ and $\phi_D$ are scalar fields that are odd under $P_{l_l}$, $P_{l_r}$ and $P_D$
respectively, and are assumed to get \emph{vevs} of order the electroweak scale, leading to:
\begin{eqnarray}
    \delta'_{i}\sim c_{4i}\frac{v^3}{M_{UV}^2m_L};\;\;
    \delta_{ij}\sim c_{ij}\frac{v^3}{M_{UV}^3}
\end{eqnarray}
$\delta'$ values with the correct magnitude are naturally obtained for $M_{UV} \approx M_{GUT}$. 
For singlet DM $\chi \approx N$, the dominant decay is through the yukawa couplings in (\ref{eq:LLN}) in which
$N$ decays to a leptonic higgs and the dark doublet $L,L^c$ followed by the decay of $L,L^c$ by mixing with the
SM leptons $L_i$ through the mixing term in (\ref{eq:LLNdecay}). Thus, the decay modes in this case are given by:
\begin{eqnarray}
    \chi &\rightarrow & (A,H)+\nu_l \rightarrow \tau^+\tau^- \nu_l
    \nonumber\\
    &or&\nonumber\\
    \chi&\rightarrow & H^{\pm}+l^{\mp} \rightarrow \tau^{\pm}l^{\mp}
    \nu_{\tau}
\end{eqnarray}
where $l=e,\mu,\tau$. This leads to different predictions for
the electron and positron spectra for different $l$. As discussed in
section \ref{positrons}, this channel will give a different fit from both
4$\tau$ and 2$\tau$ cases. In general, the spectra is expected to be a
\emph{weighted} average of the 4$\tau$, 2$\tau$, 2$\mu$ and 2$e$ cases depending on the
relative weight of channels with different $l$. This will change the fit
to PAMELA and ATIC and also the predictions for photon and neutrino fluxes. We leave the
detailed study of this case for future work.

If the DM particle is a doublet, its dominant decay mode is by mixing with the SM lepton doublet $L_i$.
There is no constraint from any flavor experiments since the operators are suppressed by the GUT scale.
So all SM leptons $L_i$ are equally likely to mix with the dark doublet. The possible decay channels
in this case are $\chi\rightarrow Z\nu_i,
~W^{\pm}l^{\mp},~H_l^{\pm}\tau^{\mp}$. Since the vector boson
channels are not only not suppressed by $\tan\beta$ but also enhanced by
$(m_{\chi}/m_W)^2$ for heavy DM (due to dominant longitudinal mode couplings at high energies),
these channels always dominate even for very small $\tan \beta$ (corresponding to $y_{\tau}\approx 1$).
This case is close to the model recently proposed in \cite{weiner-L}. However, since this case
does not fit in our original framework of the DM sector coupling to the visible sector through a leptonic higgs,
we do not discuss this case in section \ref{astro}.

Finally, in order to suppress terms like:
\begin{equation}
   \mathcal{L} \supset (H_q^{\dagger}L_iN + H_q^{T}Le_i^c)
\end{equation}
which makes the DM particles decay dominantly to $l\bar{q}q$, we also assume
that there is a quark parity $P_q$ in the quark-$H_q$ sector. If
$P_q$ is exact, this operator is completely forbidden. However,
the $\mu^2$ term in the potential (\ref{eq:DeltaV}) breaks $P_q$.
In order to simultaneously generate the $\mu^2$ term and
suppress the above operator in a consistent way, we have to break
the parity spontaneously by $\sim$ electroweak scale vevs. One way to do this is by introducing a
scalar $\phi_q$ that is odd under $P_q$ and $P_{l_l}$. So the
above operator is suppressed by
$\frac{\phi_l\phi_q\phi_D}{M_{UV}^3}$ and is suppressed compared to the
dominant decay mode (the fourth term) in (\ref{eq:LLNdecay}).
$\mu^2$ of the correct magnitude is generated if
$\mu^2=\epsilon\langle\phi_r\rangle\langle\phi_q\rangle$, where $\epsilon$ is
a technically natural small coefficient.

\subsection{Singlet Scalar DM Model}

A particularly simple form of dark matter is a singlet complex
scalar field $\Phi$. The couplings of $\Phi$ to the SM is
trivially constrained to the higgs sector, the extended higgs
sector in our picture.
With a discrete symmetry $\Phi \rightarrow - \Phi$, the most general
addition to the scalar potential is
\begin{equation}\label{eq:singlet}
    \Delta{V}~=~m_{\Phi}^2 \, \Phi^\dagger \Phi + \lambda
\Phi^\dagger \Phi (H_l^\dagger H_l +x H_q^\dagger H_q) + \lambda'(\Phi^\dagger \Phi)^2.
\end{equation}
We assume the relative strength $x$ of the quartic couplings
to $H_q$ and $H_l$ is small, i.e. $x < 1/3$, so that dark matter
annihilates dominantly via $H_l^{\dagger}H_l$ to 4 $\tau$.
Taking $m_\Phi = 4$ TeV, to allow for an explanation of the ATIC as well
the PAMELA data, we find a galactic annihilation cross section relative
to $\langle \sigma v\rangle_{std} = 3 \times 10^{-26} \mbox{cm}^3 \mbox{s}^{-1}$ by
a factor $B_{\sigma v} \approx
50 \, n^2 R$, where $\lambda = n \pi^2$ and $R$ is the Sommerfeld enhancement
factor. An attractive yukawa potential is generated between the annihilating
$\Phi$ particles by the exchange of the scalars in both $H_l$ and $H_q$, with a strength proportional
to $\lambda^2 v_l^2$ and $x^2 \lambda^2 v_q^2$, respectively. Since the dark
matter particle is expected to be about 30 times heavier than the exchanged
scalar, the resulting Sommerfeld enhancement factor can be significant.  For example,
$n=4$, and either $x \sim 1/3$ or $v_l/v_q \sim 1/3$, leads to a region where $R$
is rapidly varying from 10 to in excess of 100.  Hence $B_{\sigma v}$ of order
$10^3-10^4$, as required to explain PAMELA and ATIC, is possible, even for perturbative values of $\lambda$.
As $\lambda$ is increased, the one-loop radiative correction to the
mass term $H_l^\dagger H_l$ also increases,
leading to a little hierarchy problem of why the $H$ and $A$ states
are significantly lighter than the dark matter $\Phi$.
This is a general naturalness problem for models with annihilation of
heavy dark matter to leptonic Higgs states
lighter than 2$M_W$.   The large coupling needed for a large
annihilation cross section leads to a large radiative contribution to
the leptonic Higgs mass parameter from a loop with internal dark
matter particles.

We can also have a decaying scalar dark matter by introducing a $Z_6$ parity
so that the leading interaction that couples $\Phi$ linearly to SM particles is
the dimension 6 interaction
\begin{equation}
    \Delta{\cal L}~=~ \frac{1}{m_{UV}^2} \phi_D^3 \Phi H_l^\dagger H_l + h.c.
\end{equation}
Under the $Z_6$ symmetry, $\Phi \rightarrow - \Phi$ and  $\phi_D\rightarrow
e^{i\frac{\pi}{3}} \phi_D$, while all SM particles transform trivially.
Taking the scalar $\phi_D$ to acquire a weak scale vev, and taking $m_{UV}$
of order $10^{16}$ GeV, leads to decays of $\Phi$ to $4\tau$ with a lifetime of
order $10^{26}$ seconds. This case would require $m_{\chi}$ to be around 8 TeV (see section
\ref{astro}).


\subsection{Inert Higgs-Doublet DM Model}

Finally, we consider the scalar DM to be an electroweak doublet. A simple example of this
is the inert Higgs $H_I$, first proposed in \cite{Inert}. Extending this model, we
have three Higgs-doublets, $H_q$, $H_l$ and $H_I$, where both
$H_q$ and $H_l$ get a vev but the inert Higgs $H_I$ does not. The Higgs doublets 
are all identical in the sense of gauge charges, but have different 
masses and parities. The couplings of $H_I$ to fermions are
forbidden by dark parity. Since no symmetry can
forbid the term $(H_I^{\dagger}H_I)(H_q^{\dagger}H_q)$, we have no
symmetry explanation of suppressing the couplings of $H_I$ to the quark sector in this
model. One can only claim that it is not unreasonable to have
such couplings being numerically suppressed. However, such a possibility is at least naturally
allowed within the framework of a leptonic higgs sector.

From now on, we concentrate on couplings to leptonic higgs. The relevant new terms
in the scalar potential are:
\begin{equation}\label{Innert}
    \Delta{V}~=~m_I^2 H_I^\dagger H_I + H_I^\dagger H_I H_l^\dagger H_l
-(H_I^{\dagger}H_l)(H_l^{\dagger}H_I)+(H_I^{\dagger}H_l)^2+c.c.
\end{equation}
We omit the ${\cal O}(1)$ coefficients in front of each quartic term to
avoid use of new notation. These terms allow the lightest $P_D$ odd
particle to be neutral after electroweak symmetry breaking splits the
doublet. The second quartic term splits the charged and neutral
components of $H_I$, while the last term in (\ref{Innert}) splits
the scalar and the pseudoscalar parts of the neutral component in $H_I$, 
naturally evading bounds from direct detection. The DM particles can
annihilate to $H_l^{\dagger}H_l \rightarrow 4\tau$. However, similar to the
doublet DM in the $LL^c N$ model, the $ZZ$
annihilation channel is also present. In order to fit the ATIC data we need some
of these quartic couplings to be large and the mass splitting between
the scalar and pseudoscalar (and charged and neutral) components to be small 
(to provide a modest Sommerfeld enhancement), requiring a hierarchy between these quartic couplings.

A decaying scalar dark matter is also possible by introducing
small dark parity breaking term. Similar to the $LL^c N$ model, the lowest
dimensional operator is the mass mixing term:
\begin{equation}
    \Delta{\cal L}_{decay}~=~\delta m^2H^{\dagger}_IH_l+c.c.
\end{equation}
The corresponding term involving the quark Higgs can be forbidden
by imposing an exact quark parity as we did for the
the $LL^c N$ model. The desired small value for $\delta m^2$
can again be obtained by extending $P_l$ to the full chiral lepton parity and
the dark parity $P_D$ to $Z_4$ in which $H_I$ has 2 units of charge. Then,
the parity preserving non-renormalizable operator is:
\begin{equation}
    \Delta{\cal L}_{decay}~=~\frac{\phi^2_D\phi_l\phi_r}{m_{UV}^2}H^{\dagger}_IH_l+c.c.
\end{equation}
The parities of these $\phi$'s are as defined in the $LL^c N$ model
except the $\phi_D\rightarrow i \phi_D$ under the $Z_4$ discrete
dark symmetry. The possible decay channels are $WW$, $ZZ$,
$(A,H)\,h$, $\bar{\tau}\tau$ and the three-body decay to $(A,H)\,hh$. The
vector boson channels are enhanced by $(m_{\chi}/m_W)^2$ as before
but unlike the case for fermionic DM, these vector boson channels are also
suppressed by $\sin^2\beta$. Therefore, the $\bar{\tau}\tau$
channel can dominate when $\sin\beta$ is
relatively small. For example, for $\sin\beta\sim 10^{-2}$ ($y_{\tau}\sim 1$),
the decay branching ratio to vector bosons is about $5\%$ for 4 TeV dark matter. The scalar decay
channels come from quartic interactions (for example $\lambda_3|H_l|^2|H_q|^2$) in the potential which
can not be forbidden by any symmetry. However, the two-body decay mode is
suppressed by at least $(v_q/M_{\chi})^2\sim 10^{-3}$ while the three-body decay
mode is suppressed by an extra phase factor of $\sim 1/2\pi^2$. For quartic
couplings not much bigger than unity, the $\bar{\tau}\tau $ channel can
dominate the decay. Since the final state is $\bar{\tau}\tau$, a 4 TeV
decaying dark matter of this model is expected to fit the data well.

\section{Conclusions and Summary}\label{conclude}

In this paper we have explored the consequences of assuming that a leptonic Higgs mediates the
interactions between the dark matter sector and the Standard Model sector.  The motivation for
this approach is two-fold.  One is theoretical: the same TeV mass scale underlies both the
breaking of weak interactions and the dark matter annihilation rate, indicating a close
connection between the dark matter and Higgs sectors.  The observational motivation is that the cosmic
ray signals are {\it leptonic}; in particular, the limits on a primary $\bar{p}$ flux indicate that
mediation via a Higgs with quark couplings should be sub-dominant.  We study the minimal case
of a single leptonic Higgs, in which case its largest couplings to matter must be to $\tau$
leptons, with couplings to $e$ and $\mu$ that are sufficiently small to play no role.
The Higgs potential could lead to mass mixing between the states with
leptonic and quark couplings but, again, the absence of an exotic primary cosmic ray $\bar{p}$
flux limits this mixing to be small. Hence, there is one neutral mass eigenstate scalar, $H$,
one pseudoscalar, $A$, and one charged scalar $H^+$ that maintain their dominant leptonic couplings.
This allows an important connection between the leptonic cosmic ray signals and the expected Higgs
signatures at the LHC.  In both cases there are signals that arise from the production of
$H,A$ and $H^+$ and their subsequent decays to $\tau$ leptons.  From LEP limits and the leptonic cosmic
ray signals we argue that the states $H,A$ and $H^+$ are likely to have masses in the range of roughly
100 GeV to 200 GeV.

The leptonic cosmic ray signals, for both PAMELA and ATIC, could arise from a variety of channels:
dark matter annihilation to $\tau^4, \tau^2 \nu^2$ or dark matter decay to $\tau^4,
\tau^2, \tau^2 \nu, \tau \nu l$ via intermediate $H,A$ and $H^+$ states. For concreteness we have
computed the cosmic ray signals for the case of annihilations $\chi \chi \rightarrow \tau^4$,
and the results are shown in Figure \ref{pamatic}. Good fits to the PAMELA and ATIC data can be simultaneously
obtained for $m_\chi$ in the region around 4 TeV.  We have also commented on all the other possible 
annihilation and decay modes. The width of the peak in the leptonic signal
around 600 GeV is larger than in the case of annihilations or decays directly to $e$ and $\mu$,
and we expect this to be a common feature of all modes involving $\tau$s.  Similarly we expect
the signal in the positron fraction to continue to energies larger than the present
PAMELA data, and not to show a sharp peak.  Such an annihilation signal requires a large total boost
factor in the annihilation cross section of order $10^4$, but essentially identical cosmic ray signals
can result from dark matter decays with lifetimes of order $10^{26}$ seconds.

We have computed the photon spectrum that results from annihilations of DM via pairs of leptonic
Higgs states to $\tau^4$ in the dwarf galaxy Sagittarius. For a DM mass of 4 TeV, HESS data places
a limit of $\sim10^3$ ($\sim 10^4$) on the relevant boost factor, $B^\gamma_{tot}$, for the case that Sagittarius
has an NFW (Large Core) profile. For this annihilation mode and DM mass, the PAMELA and ATIC signals require
a boost factor $B^e_{tot}$ that is of order $10^4$.  However, these two boost factors are not identical in general,
so further observations of hard gamma rays from Sagittarius would provide a powerful probe of this
annihilation channel.  We also compute the high energy gamma ray signal expected from
DM annihilations in the center of the Milky Way galaxy, as shown in the top plot of Figure \ref{future}, which
could be detected by VERITAS 4.  The annihilation channel to $4 \tau$ also produces a galactic flux
of neutrinos which is in mild conflict with the bounds set by SuperK on upward going neutrinos
through the Earth, if the relevant neutrino and electron boost factors are identical.
As before, these boost factors can differ, so future neutrino measurements also offer the possibility
of a crucial independent verification of the DM origin of the charged lepton cosmic ray signals. In addition, 
an observation of energetic cosmic neutrino flux strongly favors the dark matter interpretation of the cosmic-ray signals
over astrophysical ones (such as pulsars) in general since astrophysical sources do not emit a large flux of high energy neutrinos.

For the decay modes, constraints from current gamma-ray and neutrino observations are easily satisfied but at the same time these also lead to less promising prospects for future experiments. For heavy decaying dark matter, neutrinos provide a better opportunity for future experiments compared to gamma rays, since the neutrino flux is proportional to the dark matter mass.

A common feaure of the cosmic ray signals discussed above is that the dark matter particle has a
mass in excess of 1 TeV and is not expected to be produced at the LHC. While cosmic ray photon and neutrino
fluxes allow discrimination between DM annihilation and decay, the LHC probes a completely complementary
feature of the theory -- the production and decay of the leptonic Higgs states $H,A$ and $H^+$. There is a very
rich and distinctive Higgs phenomenology; all the couplings of these states to matter and to
electroweak gauge bosons are determined by just their masses and two mixing angles, as shown in
(\ref{eq:SAhcouplings}).  The $\tau^4$ signal for $H\,A$ production via a virtual $Z$ is particularly
robust because it is insensitive to the values of the small Higgs mixing angle $\alpha$
and to the ratio of vevs $\tan \beta$.  Depending on the $H$ and $A$ masses this $H\,A$ production
cross section is typically in the (50-250) fb region, and from the cosmic ray signals we know that
the branching ratio to $\tau^4$ is close to unity. Detailed simulation studies are needed to estimate the 
observability of this $\tau^4$ signal.
 
We have computed  $\sigma\times$BR for the production of both the leptonic Higgs scalar $H$ and the Higgs
boson $h$ in the vector-boson ($WW$) fusion at the LHC followed by decay to $\tau$ pairs. For $m_H, m_h \lsim 2 M_W$,
discovery at the LHC is possible for a wide range of $\alpha$ and $\beta$, provided $|\alpha| > \beta$,
via the channel $\bar{\tau} \tau \rightarrow lj$, and for some regions of parameter space the signal is very large
allowing prompt discovery at the LHC.  If
$m_h > 2 m_H$ or $2 m_A$ the $\tau^2$ signal for $h$ production is lost, but the cascade
$h \rightarrow HH,AA \rightarrow  \tau^4$ leads to a 4$\tau$ signal instead.  Comparisons with benchmark studies
for $4 \tau$ signals in the next-to-minimal supersymmetric theory suggest that a 5$\sigma$ discovery may
be possible with an integrated luminosity of 30 $fb^{-1}$ for $m_h < 130$ GeV.  Further simulation
studies are needed, especially for $m_h > 130$ GeV where backgrounds from top quark pair production become
important. Other LHC Higgs signals, such as charged higgs pair-production by Drell-Yan followed by their decay to 
$\tau ^{\pm}\nu$, and a spectacular 8$\tau$ signal from $h$ pair-production, are also possible and would be quite 
interesting to study further.

Simple particle physics models of DM that couples to the visible sector via a leptonic Higgs are
very easy to write down.  In the case that DM is a heavy lepton, the interaction with the leptonic
Higgs is via a yukawa coupling, allowing $s$-wave annihilations to $H\,A$.  On the other hand,
scalar dark matter can annihilate via a quartic scalar interaction to $H$ or $A$ pairs.
In either case a large annihilation cross section is possible, aided in many cases by a
modest Sommerfeld boost factor; but in both cases the DM abundance must be produced non-thermally.  Alternatively,
the DM could be produced thermally, with the cosmic ray signals arising from decays. The long lifetime
follows from a decay amplitude that is suppressed by two powers of the ratio of the weak scale to the
unified scale, which arises from a combination of breaking the discrete symmetry that leads to near
stability of the DM and the discrete symmetry that ensures the leptonic nature of the Higgs.

\acknowledgments{We would like to thank Yasunori Nomura, Alessandro Strumia and Jesse Thaler for useful discussions.
HG would like to thank the KITPC for their hospitality where part of his research was conducted.
This work is supported by the U.S. Department of Energy under contract no. DE-AC02-
05CH11231 and NSF grant PHY-04-57315 .}

\end{document}